\documentclass[12pt]{JHEP3}
\usepackage{amsfonts}
\usepackage{epsfig}
\def\cF{{\cal F}}

\def\Dsl{\hbox{/\kern-.6700em\it D}} 
\def\dsl{\hbox{/\kern-.5300em$\partial$}}

\def\eqa{\begin{eqnarray}}
\def\eeqa{\end{eqnarray}}
\def\eeq{\end{equation}}

\def\pref#1{(\ref{#1})}
\def\td{{\tilde{d}}}
\def\tgamma{{\tilde{\gamma}}}
\def\hgamma{{\hat{\gamma}}}
\def\ls{{l_s}}
\def\dperp{d_{T}}
\def\roughly#1{\mathrel{\raise.3ex\hbox{$#1$\kern-.95em
\lower1ex\hbox{$\sim$}}}}
\def\lsim{\roughly<}

\title{Branonium}
\author{C.P. Burgess,$^1$, P. Martineau,$^1$ F. Quevedo$^2$
 and R. Rabad\'an$^3$
\\

$^1$ Physics Department, McGill University,  3600 University Street,
 Montr\'eal, Qu\'ebec, Canada, H3A 2T8.\\

$^2$ Centre for Mathematical Sciences, DAMTP,
               University of Cambridge,\\
               Cambridge CB3 0WA UK.\\

$^3$ Theory Division CERN, CH-1211 Gen\`eve 23, Switzerland
}

\abstract{ We study the bound states of brane/antibrane systems by
examining the motion of a probe antibrane moving in the background
fields of $N$ source branes. The classical system resembles the
point-particle central force problem, and the orbits can be solved
by quadrature. Generically the antibrane has orbits which are not
closed on themselves. An important special case occurs for some
D$p$-branes moving in three transverse dimensions, in which case
the orbits may be obtained in closed form, giving the standard
conic sections but with a nonstandard time evolution along the
orbit. Somewhat surprisingly, in this case the resulting
elliptical orbits are {\it exact} solutions, and do not simply
apply in the limit of asymptotically-large separation or
non-relativistic velocities. The orbits eventually decay through
the radiation of massless modes into the bulk and onto the branes,
and we estimate this decay time. Applications of these orbits to
cosmology are discussed in a companion paper.}

 \preprint{McGill-03/04}

\newcommand{\sfrac}[2]{{\textstyle\frac{#1}{#2}}}

\def\be{\begin{equation}}
\def\ee{\end{equation}}
\def\bea{\begin{eqnarray}}
\def\eea{\end{eqnarray}}

\begin{document}

\section{Introduction}
One of the most important developments in string theory of late
has been the realization that it contains many kinds of extended
objects like D$p$-branes. In many situations these objects can be
treated much like elementary particles, in that they carry a
generalised mass and charge and these give rise to mutual forces
which govern their relative motion.

This observation is all the more interesting when the dimension of
the brane exceeds three, since then the moving branes can contain
an entire world inside of themselves (possibly including us). In
this kind of picture the collision of D$p$-branes can be
interpreted as colliding universes. This observation has recently
been used to propose several novel scenarios for the very-early
universe, giving rise to inflation
or alternatives to inflation
\cite{BI,OurBI,dvali,GRZ,BIO,majumdar,Ekpyrosis,cyclic,rev},
 allowing in both
cases an attractive geometrical and stringy interpretation of the
various cosmologically-active scalar fields.

Perhaps the best-studied brane interactions are brane collisions,
which in the particular case of branes and antibranes can lead to
brane annihilation (for which  insights have come from
recent studies of tachyon condensation). However these kinds of
direct collisions only begin to explore the rich kinds of physics
which are possible when particles and branes mutually interact. A
different --- and for D$p$-branes as yet largely unexplored ---
possibility is the development of bound orbits, such as commonly
arise for particles interacting through central forces.

In this paper we begin the study of the properties of the brane
generalisation of such bound orbits. We do so partly motivated by
the potential applications of such a study to brane-world
cosmology. In particular we could imagine the brane which makes up
our own universe being in a closed orbit around another brane due
to the mutually attractive forces which act between them. But
regardless of these applications, we regard brane bound states to
be of interest in their own right inasmuch as they add to our
understanding of how branes and strings can interact.

We start this study here with a discussion of the bound states
which are possible for brane/antibrane systems. These are among
the simplest to use in applications, partly because of the
triviality of the various consistency conditions which would
otherwise arise when trying to construct brane configurations
inside compact extra dimensions. We call this kind of bound state
`branonium' due to its obvious similarity with the analogous
electron/positron system: positronium.

Our discussion is couched in terms of a test antibrane which is
moving within the fields sourced by a system of $N$ parallel
branes. We set up the effective brane lagrangian for this motion
and analyse the features of the resulting bound orbits. It happens
that despite the many complications of the resulting effective
Lagrangian, it is still possible to solve explicitly for the
trajectory of the extended test brane, with a result which is very
similar to those obtained for nonrelativistic motion in a central
force. In particular we find the conditions for the existence of
stable closed orbits, and show that these include nontrivial
systems of D-branes, with perfectly elliptical (non-precessing)
orbits being possible if the branes move in three transverse
dimensions.

Finally, we give a preliminary study of the stability of the
system due to different sources, including radiation reaction due
to radiation of bulk and brane modes, as well as nonperturbative
instabilities and tachyonic modes.
The dominant instability appears to be towards radiation into
massless modes, and we compute the orbital decay time for this
process.

\section{Probe-Brane Analysis}
We start by setting up and solving the equations which are
relevant to the motion of a probe antibrane moving within the
background of $N$ parallel branes. To do so we first examine the
fields which the $N$ source branes set up, and then examine the
equations of motion which they imply for a test brane or
antibrane. We show how these equations may be solved for the
motion of the test-brane centre of mass when all other brane modes
are frozen. (We return to the question of {\it why} these modes
should be frozen in the next section.) Finally we examine the
bound-state orbits and identify the surprising special class for
which the exact {\it relativistic} orbits are ordinary conic
sections, and we study the properties of the orbits in this
special case in some detail.

\subsection{The Fields Due to $N$ Source Branes}
Let us first consider the gravitational, dilaton and
`electromagnetic' fields which are set up by a set of $N$ parallel
$p$-branes in $D$-dimensional spacetime. We take the action for
these fields to be given in the Einstein Frame by
\begin{equation}
    \label{eq:EFbulkaction}
    S_s = - \int d^Dx \, \sqrt{- g} \left[ \frac12 \, g^{MN} \Bigl(
    R_{MN} + \partial_M \, \phi \, \partial_N \, \phi \Bigr) + {1
    \over 2 \, n!} \;e^{\alpha \phi} \, F_{M_1...M_n} \, F^{M_1...M_n}
    \right]\,,
\end{equation}
where the $n$-form field strength is related to its $(n-1)$-form
gauge potential in the usual way $F_{[n]} = dA_{[n-1]}$. Here $n$
is related to the spacetime dimension $d$ of the $N$ $p$-branes by
$d = p + 1 = n - 1$. We denote the $d$ coordinates parallel to the
branes by $x^\mu$ and the $D-d$ transverse coordinates by $y^m$.

The constant $\alpha$ depends on which kind of brane is being
considered. For instance, if $A_{[n-1]}$ were to arise from a
string-frame action
\begin{equation}
    \label{eq:saction}
    S_{\rm SF} = - \, {1 \over 2 \, n!} \int d^Dx \,
    \sqrt{- \hat{g}} \;e^{\alpha_s \phi}
    \, F_{\hat{M}_1...\hat{M}_n} \, F^{\hat{M}_1...\hat{M}_n} \, ,
\end{equation}
with the string-frame metric given by $\hat{g}_{MN} = e^{\lambda
\phi} \, g_{MN}$ for $\lambda = 4/(D-2)$, then $\alpha = \alpha_s
+ 2 - \lambda(n-1)$. Two cases of particular interest are: ($i$)
NS-NS fields (like the fields rising in the gravity supermultiplet
in various dimensions), for which $\alpha_s = -2$ and so $\alpha =
\alpha_{NS} = -4(n-1)/(D-2)$; ($ii$) R-R fields, for which
$\alpha_s = 0$ and so $\alpha = \alpha_R = 2(D-2n)/(D-2)$. In the
special case $D=10$ we have $\lambda = \sfrac12$ and so these two
cases become $\alpha_{NS} = -\, \sfrac12 (n-1) = -\,\sfrac12(p+1)$
or $\alpha_R = \sfrac12(5-n) = \sfrac12(3-p)$, respectively.

The solutions to the equations for this system which describe the
fields due to a stack of $N$ parallel collocated $p$-branes is (in
the Einstein Frame) \cite{Kelly}:
\be \label{Dbrane}
    ds^2 = h^{-\tgamma} dx^2 +  h^{\gamma} dy^2 \, ,\qquad e^{\phi} =
    h^{\kappa} \, ,\qquad A_{01...p} = \zeta (1 -  h^{-1}) \, ,
\ee
where all the other components of the $(n-1)$-form field vanish.
The constants $\gamma$, $\tgamma$, $\kappa$ and $\xi$ are given by
\be \label{braneconsts}
    \gamma = {4 d \over (D-2) \, \Delta} \, , \qquad
    \tgamma = {4 \td \over (D-2) \, \Delta} \, , \qquad
    \kappa = {2 \alpha \over \Delta} \, , \qquad
    \zeta = {2 \over \sqrt\Delta} \, ,
\ee
where  $\td = D - d - 2$ and
\be \label{Deltadef}
    \Delta = \alpha^2 + \frac{2 \, d \, \td}{(D-2)} \, .
\ee
We assume throughout the existence of more than 2 transverse
dimensions, so $\td > 0$. The quantity $h$ denotes the harmonic
function:
\begin{equation} \label{harmonic}
    h(r) = 1 +  k/r^{\td} \, ,
\end{equation}
where $r$ is the radial coordinate transverse to the branes, $r^2
= \delta_{mn} \, y^m y^n$, and $k$ is an integration constant.

D$p$-branes in 10 dimensions represent a particularly interesting
special case of this solution, corresponding
\cite{JoesBigBookoString} to the choices $\alpha = \alpha_R =
\sfrac12(3-p)$, $\Delta = 4$, $d = p+1$, $\td = 7-p$, $\kappa =
\sfrac14(3-p)$ and $k = c_p \, g_s N \, \ls^\td$. (Clearly the
condition $\td
> 0$ then implies $p < 7$.) Here $\ls$ is the string length, which
is related to the string tension by $\ls^2 = \alpha'$. The string
coupling constant at infinity is denoted $g_s$ and $c_p$ is the
constant $c_p = (2 \sqrt{\pi})^{5-p} \, \Gamma(\frac{7-p}{2})$.

Having an explicit string solution is useful for many reasons, not
the least of which is the understanding which this gives of the
limits of validity of the field theoretic description we shall use
throughout what follows. For these purposes there are two
dimensionless quantities which are relevant. The first of these
controls the validity of string perturbation theory, which for
string self-interactions requires the local string coupling,
$g(r) = g_s \, e^\phi$, to be everywhere small. The smallness of
this quantity is required in order to suppress the modifications
which higher string loops would otherwise have on our
semiclassical analysis.

The second important dimensionless quantity is the ratio $r/\ls$,
which must be large in order to believe that the brane
interactions are dominated only by the mediation of bulk fields
which correspond to massless string states. Once $r \sim \ls$
higher order corrections in $\alpha'$ become important because
over these short distances contributions from higher string modes
become possible. A particularly dangerous example of such a
correction arises if a probe antibrane is introduced, since then
there is an open string tachyon which appears at short distances
within the open-string sector for strings which stretch between
the $N$ branes and the antibrane. Within the closed string picture
this tachyon can be regarded as a kind of $\alpha'$ correction,
since by duality it can be thought of as emerging when all the
closed string modes are resummed. The appearance of this tachyon
signals the onset of the instability in which the probe antibrane
annihilates with one of the source branes.

One might worry that the requirement $r \gg \ls$ might also
preclude using the full nonlinear solution given above, requiring
us instead to use only its leading approximation in powers of
$k/r^\td$. But this need not be required if the solution's
characteristic length scale, $\rho = k^{1/\td} = \ls (c_p \, g_s
N)^{1/(7-p)}$, is also large, $\rho \gg \ls$, since in this case
we may have $k/r^\td = O(1)$ even if $r \gg \ls$. Notice that for
D$p$-branes the condition $\rho \gg \ls$ requires $g_s N \gg 1$,
and so takes us beyond the limit of open-string perturbation
theory, which requires $g_s N \ll 1$ (where the factor of $N$ here
arises from the Chan-Paton contributions from each world sheet
boundary). In this way the field-theory limit provides (for $r \gg
\ls$) results which are nonperturbative in the coupling of the
bulk fields to the branes.

\subsection{The Probe Brane}
In the above background let us now follow the motion of another
single parallel $p$-brane, displaced from the original $N$ by the
radial coordinate-distance $r$
(for a related discussion see \cite{probe}). This motion is described by the
brane action, which can be decomposed as the sum of two pieces:
the Born-Infeld (BI) and the Wess-Zumino (WZ) parts. The dilaton
and graviton couplings are given by the Born-Infeld contribution,
which is (in the String Frame):
\begin{equation} \label{BornInfeld}
    S_{BI} = -T_p \int d^d \xi \; e^{-\phi} \sqrt{-
    \det(\hat{g}_{MN} \,
    \partial_{\mu} x^M \partial_{\nu}x^N + 2\, \ls^2 \cF_{\mu \nu})}
    \, ,
\end{equation}
where $x^M$ are the coordinates of the embedding, $\xi^\mu$ are
the world-volume coordinates and $T_p$ denotes the brane tension.
$\cF_{\mu\nu}$ denotes the field strength for any open-string
gauge modes confined to the brane, although for the moment we put
these fields to zero.

The coupling to the bulk gauge field, $A_{[p+1]}$, is given by the
Wess-Zumino part of the brane action:
\begin{equation} \label{WZ}
    S_{WZ} = -q \, T_p  \int  A_{[p+1]} \, ,
\end{equation}
where $q$ represents the brane charge, with $q = 1$ representing a
probe brane and $q=-1$ representing a probe antibrane.

Since our branes are straight and parallel it is convenient to
choose the following coordinate gauge:
\begin{equation}
    x^{\mu} =  \xi^{\mu} \, ,
\end{equation}
where as before $x^{\mu}$ are the coordinates parallel to the
world volume. To the extent that we follow only the brane's
overall centre-of-mass motion --- we return to the validity of
this assumption in the next section --- we may also take
\begin{equation}
    y^{m} = y^m(t)
\end{equation}
where $t = x^0$ denotes coordinate time and $y^{m}(t)$ are the
coordinates perpendicular to the brane surface, which parameterise
the position of the probe. With these choices the induced
String-Frame metric on the probe becomes:
\begin{equation}
    d\hat{s}^2 = e^{\lambda \phi} \, ds^2 =  h^{-\beta} \left[ \left(
    - 1 + h^\omega v^2 \right) dt^2 + (d\xi^{i})^2\right]
\end{equation}
where $i = 1,...,p$ and $v^2 = \sum_m ({dy^m}/{dt})^2$. The
constants $\beta$ and $\omega$ are given by
\be
    \beta = \tgamma - \kappa \lambda = \frac{4 (\td - 2
    \alpha)}{(D-2) \, \Delta} \, , \qquad
    \omega = \gamma + \tgamma = \frac{4}{\Delta} \, .
\ee
For the special case of D$p$-branes in 10 dimensions these become
$\beta = \frac12$ and $\omega = 1$.

The effective action for the probe brane is then:
\begin{equation} \label{FullAction}
    S = - T_p \, V_p \int dt \; \left[ \frac{1}{h^{\eta}} \sqrt{1 - h^\omega \, v^2}
    +  q \, \zeta \left( 1 - \frac{1}{h} \right) \right]
\end{equation}
where $V_p = \int d^p\xi$ is the spatial volume spanned by the
probe brane, $\zeta = \sqrt\omega$ given by eq.~\pref{braneconsts}
and
\be
    \eta = \kappa + \frac{\beta \, d}{2} = \frac{2 \, [ \alpha(\td
    - d) + d \, \td]}{(D-2) \, \Delta} \, .
\ee
For the special case of D$p$-branes in 10 dimensions we therefore
have  $\eta = 1$.

Apart from an additive constant, the brane action has the form
\begin{equation}
    S = - m \int dt \; \left[ \frac{1}{h^{\eta}}
    \sqrt{1 - h^\omega \, v^2} - \frac{\hat{q}}{h} \right]
\end{equation}
with $\hat{q} = \zeta \, q$ and $m = T_p V_p$. This is very
similar to that of a particle moving in the $D-d$ dimensions
transverse to the branes. That part of the action containing the
square root corresponds to the particle's relativistic kinetic
energy and the coupling to the dilaton field, while the part
involving $\hat{q}$ corresponds to the potential due to the
Ramond-Ramond couplings. Notice that the apparent mass, $m$, of
the particle scales with the brane world-volume, and so becomes
infinitely large in the limit that this volume diverges.

Several features of the probe-brane action bear emphasis:
\begin{itemize}
\item A measure of the potential energy of the probe brane may be
obtained by evaluating the brane action for a brane at rest:
$v=0$. The result obtained in this way is proportional to the
combination $h^{-\eta} - \hat{q}/h$. For D$p$-branes in 10
dimensions this becomes $(1-q)/h$, which is nonzero for a probe
antibrane ($q=-1$), but vanishes for probe brane ($q=1$). This is
as one expects given the BPS nature of the D$p$-brane.
\item The brane motion is along a null geodesic in the transverse dimensions
when $v^2 = h^{-\omega}$. This becomes $v = 1$ asymptotically far
from the set of N source branes, but is smaller than unity closer
to the source branes.
\item
Massless particles trapped on the probe brane move along null
geodesics of the induced metric, which corresponds to
$(d\xi^{i}/dt)^2  = 1 - h^\omega v^2$, indicating that the
effective maximum speed for trapped particles, $c^2 = 1 - h^\omega
v^2$, varies with time as the brane moves through the fields of
the source branes. Notice that $c^2 \to 0$ if the brane becomes
ultra-relativistic: $v^2 \to h^{-\omega}$.
\end{itemize}

\subsection{Classical Dynamics}
Let us analyse the movement in $\dperp = D- d$ dimensions of a
particle with the Lagrangian:
\begin{equation} \label{ParticleL}
L = -  m \left( \frac{1}{h^{\eta}} \sqrt{1 - h^\omega \, v^2} -
\frac{\hat{q}}{h}\right) \, ,
\end{equation}
with $h = 1 + k/r^\td$ and $k > 0$.

Invariance of this Lagrangian under $O(\dperp)$ rotations ensures
the conservation of the angular momentum tensor, $L_{ij} = x_i \,
p_j - x_j \, p_i$, with $x_i$ and $p_j$ denoting the components of
the particle's position and canonical momentum vectors. The
conservation of this tensor ensures that the particle motion is
confined to the plane which is spanned by the particle's initial
position and momentum vectors.
Let us denote by $r$ and $\theta$ polar coordinates in the plane
of the particle motion. The Lagrangian for the resulting
two-dimensional system is
\begin{equation}
L = -  m \left( h^{-\eta} \sqrt{1 - h^\omega \, (\dot{r}^2 + r^2
\dot{\theta}^2)} - \hat{q} \, h^{-1} \right) \, .
\end{equation}

Invariance under translations of $\theta$ ensures the conservation
of angular momentum
\begin{equation}
\ell = \frac{p_{\theta}}{m}  = \frac{r^2 \dot{\theta}}{ h^{\eta -
\omega} \sqrt{1 - h^\omega \, (\dot{r}^2 + r^2 \dot{\theta}^2)}}
\end{equation}
Notice that this is just the relativistic angular momentum. In
terms of this the angular velocity becomes
\begin{equation} \label{angvelocity}
r^2  \dot{\theta}^2 = \frac{\ell^2 \, h^{2(\eta - \omega)}(1 -
h^\omega \, \dot{r}^2)}{r^2 + \ell^2 \, h^{2\eta - \omega}} \, .
\end{equation}

The canonical momentum associated with the radial motion is
similarly
\begin{eqnarray}
    \rho_r = \frac{p_{r}}{m}  &=&  \frac{\dot{r}}{h^{\eta - \omega} \,
    \sqrt{1 - h^\omega \, (\dot{r}^2 + r^2 \dot{\theta}^2)}} \, ,
    \nonumber \\
    &=& \frac{\dot{r}}{h^{\eta - \omega}} \; \left( \frac{1 +
    \ell^2 \, h^{2 \eta - \omega} /r^2}{1 - h^\omega \, \dot{r}^2}
    \right)^{1/2} \, .
\end{eqnarray}
Inverting this to obtain $\dot{r}$ as a function of $\rho_r$ then
gives
\be
    \dot{r}^2 = \frac{h^{2(\eta - \omega)} \, \rho_r^2}{ 1 +
    h^{2\eta - \omega} \left( \ell^2/r^2 + \rho_r^2 \right)} \, .
\ee

Constructing the canonical Hamiltonian gives the energy formula:
\begin{equation} \label{energia3}
    \varepsilon = \frac{E}{m} = \frac{1}{h^\eta}
    \left[ 1 + h^{2\eta - \omega} \,
    \left( \frac{\ell^2}{r^2} + \rho_r^2 \right)
    \right]^{1/2} - \frac{\hat{q}}{h}
    \, ,
\end{equation}
which is conserved during the motion.

\subsubsection{Qualitative Analysis}
A qualitative understanding of the orbits which are
implied by this system may be obtained using the effective
potential
\be \label{effpot}
    V_{\rm eff}(r) = \varepsilon(\rho_r = 0) =
    \frac{1}{h^\eta} \left[ 1 + h^{2\eta - \omega} \,
    \left( \frac{\ell^2}{r^2} \right) \right]^{1/2}
    - \frac{\hat{q}}{h} \, ,
\ee
which we plot in Figure \pref{VeffFig} for the case $p=6$.

The utility of this potential follows from the observation that
$\varepsilon$ is a monotonically increasing function of $\rho_r$:
$\partial \varepsilon/\partial \rho_r \ge 0$. The allowed range of
$r$ for classical motion is therefore easily found by the standard
device of plotting $V_{\rm eff}(r)$ against $r$, and finding those
$r$'s for which $\varepsilon \ge V_{\rm eff}(r)$.

The behaviour of $V_{\rm eff}$ follows from the properties of $h$,
which is a monotonically decreasing function of $r$, with limits
$h \to k/r^\td \to \infty$ as $r \to 0$ and $h \to 1$ as $r \to
\infty$. This implies
\be \label{Veffrtoinfty}
    V_{\rm eff} \to (1-\hat{q}) + \cases{(\hat{q} - \eta)k/r^\td &
    for $\td < 2$\cr
    [\ell^2/2 + \hat{q} - \eta]k/r^2 & for $\td = 2$\cr \ell^2/(2 r^2)
    & for $\td > 2$ \cr} \qquad \hbox{as $r \to
    \infty$} \, ,
\ee
and, provided $\eta$ and $\omega$ are positive,
\be \label{Veffrto0}
    V_{\rm eff} \to \cases{\ell r^{-1+\omega \td/2}/k^{\omega/2} &
    for $(\omega - 2)\, \td < 2$\cr
    [\ell/k^{\omega/2} - \hat{q}/k]r^\td & for $(\omega - 2)\, \td = 2$\cr
    - \hat{q} r^\td/k & for
    $(\omega - 2)\, \td > 2$ \cr} \qquad \hbox{as $r \to 0$} .
\ee

These limiting forms are often sufficient to determine the
existence of a minimum for $V_{\rm eff}$ for some value $r = r_c$,
which signals the existence of a stable circular orbit at this
radius. For instance, for the 10 dimensional D$p$-brane example we
have $\omega = \eta = 1$ and $\td = 7 - p > 0$. This then gives
the limits
\bea
    V_{\rm eff} &\to& (1-q)+ \cases{(q-1)k/r &
    for $p=6$\cr
    [\ell^2/2 + q - 1]k/r^2 & for $p = 5$\cr \ell^2/(2 r^2)
    & for $p < 5$ \cr} \qquad \hbox{as $r \to
    \infty$} \, , \nonumber \\
    V_{\rm eff} &\to& \ell r^{(5-p)/2}/k^{1/2}
    \qquad \hbox{as $r \to 0$} .
\eea

\FIGURE{ \centering \epsfxsize=3in \hspace*{0in}\vspace*{.2in}
\epsffile{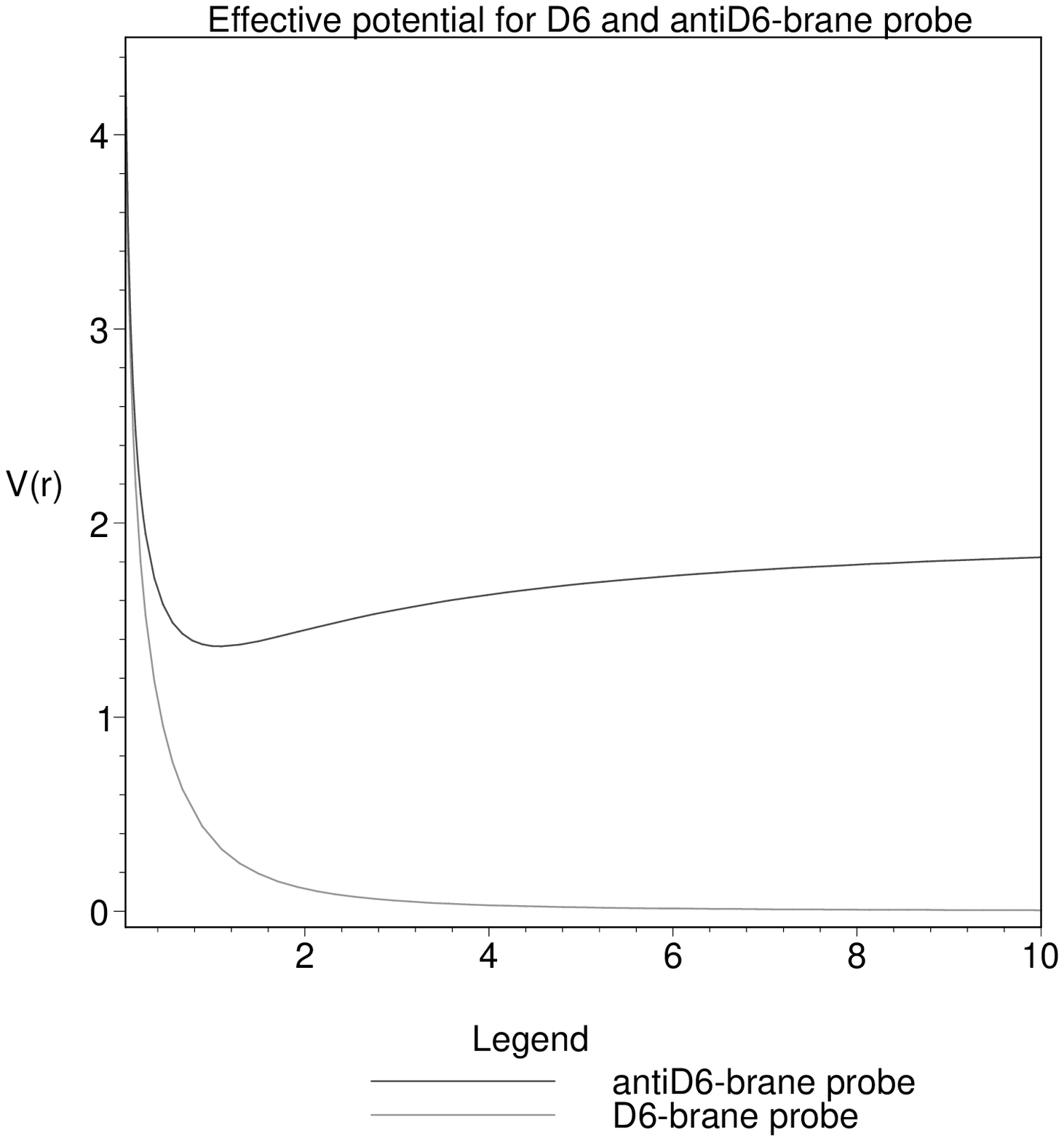} \caption{\small The effective potential
for the radial motion of the brane centre-of-mass for two
different probes, a D6 and an anti-D6-brane in the background of a
large number $N$ of D6-branes. We have taken $N g_s l_s = 2$.
Bound states exist only for anti-D6-branes.} \label{VeffFig} }

These expressions show that for $p > 5$ $V_{\rm eff}$ blows up (to
positive infinity) at the origin, and goes asymptotically to a
constant as $r \to \infty$. Furthermore, if $q = -1$ then $V_{\rm
eff}$ begins decreasing as $r$ comes in from infinity, and so
there must be a minimum somewhere in between. The position, $r_c$,
of the minimum is the solution to the equation $(2r_c +1)^2
\ell^2=8 r_c^3$.

For $p < 5$, $V_{\rm eff}$ instead approaches zero both as $r \to
0$ and as $r \to \infty$, and it grows as $r$ comes in from
infinity. The effective potential in this case instead has a
maximum at some intermediate radius, and no stable circular orbits
exist. Bound motion exists for distances smaller than the one
corresponding to the maximum of the potentials but the trajectories
are not closed and eventually pass through the origin.

The angular frequency, $\Omega$, for a circular orbit is a
constant, which is related to the orbital radius, $r_c$, by
eq.~\pref{angvelocity} specialized to $\dot{r} = 0$:
\begin{equation}
\Omega^2 = \frac{\ell^2 \, h_c^{2(\eta - \omega)}}{r_c^2 \left(
r_c^2 + \ell^2 \, h_c^{2\eta - \omega} \right)} \, ,
\end{equation}
with $h_c = h(r_c)$.

These circular orbits, when they exist, have the minimum energy
for a given angular momentum, and so are stable (at least within
the present framework). If the energy is increased slightly, for
fixed $\ell$, a broader class of orbits arises for which $\rho_r
\ne 0$ and so for which the radial coordinate oscillates between
the nearest classical turning points, $r_\pm$, defined as the two
roots (on either side of $r_c$) of the condition $\varepsilon =
V_{\rm eff}(r_\pm)$. For $p > 5$ the resulting orbits remain
bounded away from $r = 0$ since they oscillate around the radius
of the stable circular orbit. For $p < 5$ by contrast the bound
orbits have radii which range from zero up to a turning point
closer to $r = 0$ than the position of the maximum of $V_{\rm
eff}$.

In general these orbits are not closed, since the period of the
radial oscillations need not be commensurate with the angular
period. In this case the orbits will be ergodic in the sense that
the system eventually passes through all points which are allowed
by the given conserved energy and angular momentum. Since for $p <
5$ these orbits include radii near $r=0$, the probe brane can and
will pass near the source branes, and so is likely to be quickly
annihilated.

The existence of such unclosed orbits is not a surprise, given
that similar orbits also arise for solutions to the relativistic
Kepler problem (such as for the precession of the orbit of
Mercury). What we shall find which {\it is} a surprise is that for
some brane systems the orbits {\it are} closed non-precessing
ellipses, even once relativistic effects are included.

\subsubsection{Explicit Solutions}
We now turn to a more explicit characterization of the orbits.
Notice that the above equations of motion imply that the
derivative $dr/d\theta$ takes the simple form:
\be \label{rprime}
    \frac{dr}{d\theta} = \frac{\dot{r}}{\dot{\theta}} = \frac{r^2
    \, \rho_r}{\ell} \, .
\ee

To solve for the trajectory followed by the particle we start from
eq.~\pref{rprime} and use eq.~\pref{energia3} to eliminate
$\rho_r$ in favour of the conserved $\varepsilon$. Using the
standard change of variable, $u = 1/r$, we have $u' = du/d\theta =
- \rho_r/\ell$ and so
\begin{equation}
\varepsilon = \frac{1}{h^\eta} \left( \Bigl[ 1 + h^{2\eta -
\omega} \ell^2 (u^2 + u'^2) \Bigr]^{1/2} - \frac{\hat{q}}{h}
\right) \label{E}
\end{equation}
where now $h(u) = 1 + k \, u^\td$. The orbit is then obtained by
quadrature:
\begin{equation}
\theta - \theta_0 = \ell \int_{u_{0}}^u  \frac{dx}{F(x)}
\label{quadrature}
\end{equation}
where
\be \label{QIntegrand}
    F(x) = \Bigl[ \varepsilon^2 \, h^\omega(x) +
    2 \varepsilon \, \hat{q} \, h^{\omega - 1}(x) +
    \hat{q}^2 \, h^{\omega - 2}(x) - h^{\omega - 2 \eta}(x) - \ell^2 x^2
    \Bigr]^{1/2} .
\ee

\subsubsection{D$p$-branes}
Again 10 dimensional D$p$-branes provide a particularly
interesting special case, corresponding to the choices $\eta =
\omega = 1$ and $\hat{q} = q = \pm 1$. In this case the Lagrangian
\pref{ParticleL} becomes
\begin{equation} \label{grangiano}
L = - \, \frac{m}{h} \left[ \sqrt{1 - h \, (\dot{r}^2 + r^2
\dot{\theta}^2)} - q \right] \, ,
\end{equation}
and so the canonical momenta become
\begin{eqnarray}
\ell &=& \frac{r^2 \dot{\theta}}{\sqrt{1 - h \, (\dot{r}^2 + r^2
\dot{\theta}^2)}} \, , \nonumber\\
   \hbox{and}\qquad \rho_r &=& \dot{r} \; \left( \frac{1 +
    \ell^2 \, h /r^2}{1 - h \, \dot{r}^2}
    \right)^{1/2} \, .
\end{eqnarray}
The conserved energy then is
\begin{equation} \label{D6energy}
    \varepsilon = \frac{1}{h} \left\{
    \left[ 1 + h \,
    \left( \frac{\ell^2}{r^2} + \rho_r^2 \right)
    \right]^{1/2} - q \right\}
    \, .
\end{equation}

In this case eqs.~\pref{quadrature} and \pref{QIntegrand} defining
the orbits become
\begin{equation}
\theta - \theta_0 = \int_{u_{0}}^u  \frac{dx}{\sqrt{A + B \, x^\td
- x^2}} \label{teta}
\end{equation}
where $A = (\varepsilon^2+ 2 \, q \, \varepsilon)/\ell^2$ and $B =
k \, \varepsilon^2/\ell^2$. What is striking about this solution
is that it is precisely the same orbit as would be given by
classical nonrelativistic motion in the presence of a central
potential of the form $V_{c} = - m_c k_c/r^\td$:
\be \label{NRref}
    \frac{L_{c}}{m_c} = \frac{1}{2} \, \left(\dot{r}^2 + r^2 \,
    \dot{\theta}^2 \right) + \frac{k_c}{r^\td} \, .
\ee
The orbits for this Lagrangian \cite{g} are given by
eq.~\pref{teta}, but with $A = A_{c} = 2 \varepsilon_c /\ell_c^2$
and $B = B_c = 2 k_c /\ell_c^2$, where as before $\varepsilon_c =
E_c/m_c$ and $\ell_c = p_\theta/m_c$ are respectively the energy
and angular momentum per unit mass. These correspond to replacing
$k_c = 2\, k$ and taking the nonrelativistic limit: $\varepsilon =
2 + \varepsilon_c$ with $|\varepsilon_c| \ll 1$.

This shows that the orbits of the fully relativistic D$p$-brane
system have {\it exactly} the same shape as those of the
nonrelativistic system --- given by eq.~\pref{NRref} --- with the
corresponding orbital parameters read off by making the
substitution $(A_c,B_c) \leftrightarrow (A,B)$. In particular, for
bound states we take $\varepsilon_c < 0$ and so have $A_c < 0$. It
follows that the bound D$p$-brane orbits must also have $A < 0$.
Now for D$p$-branes the energy expression, eq.~\pref{D6energy},
implies $\varepsilon \ge 0$ for both $q = \pm 1$. Consequently $A
= (\varepsilon^2 + 2 \, q \, \varepsilon)/\ell^2$, can only be
negative for antibranes, since then $q = -1$. This is as expected
since only if the probe is an antibrane does it experience an
attractive net force.

It is quite remarkable that the orbits for this system are exactly
solvable at the same level as are those of the analogous
nonrelativistic central force systems, particularly given that the
D$p$-brane Lagrangian includes an infinite number of terms when
written in powers of $1/r$ and $v^2$. We note in passing that
although the shape of the orbits are unchanged from the
nonrelativistic central-force problem, the time evolution of the
branes along the orbits {\it does} change, as will be seen in more
detail in a later section.

\subsubsection{Branes at Large-Distances}
To this point we have studied the orbits using the full solutions
to the nonlinear field equations, which for D$p$-branes can be
done within a controlled approximation provided $g_s N \gg 1$. We
now turn to the case where $r$ is sufficiently large that only the
weak-field form for the fields is required. This might be expected
to apply in the case where $N$ is not large.

For large $r$ the virial theorem states that $\sfrac12 \, v^2 \sim
k/r^\td$ and so if we expand in powers of $k/r^\td$ we must also
expand to the same order in powers of $v^2$. Expanding the
Lagrangian, eq.~\pref{ParticleL}, in this way gives (up to an
additive constant)
\bea
    S &=& T_p V_p \, \int dt \left\{ \left[ (\eta -
    \hat{q}
    ) \, \frac{k}{r^\td} + \frac{v^2}{2} \right] \right. \nonumber \\
    && \qquad\qquad \left. + \left[
    \left(\hat{q} - \, \frac{\eta\, (\eta + 1)}{2} \right) \left(
    \frac{k}{r^\td} \right)^2 + (\omega - \eta) \, \frac{k \,
    v^2}{r^\td} + \, \frac{v^4}{8} \right] + \cdots \right\}\, .
\eea

For D$p$-branes this reduces to
\be
    S = T_p V_p \, \int dt \left\{ \left[ (1 -
    q) \, \frac{k}{r^\td} + \frac{v^2}{2} \right] + \left[
    \left( q - 1\right) \left(
    \frac{k}{r^\td} \right)^2 + \frac{v^4}{8} \right] + \cdots \right\}\, .
\ee
so we see that the leading dynamics is that of a particle moving
in the presence of the central potential $V = (q-1)k/r^\td$, which
is attractive for $q = -1$. This is what is expected from explicit
calculations using the string cylinder graph. The orbits for this
case are given by eq.~\pref{teta}, but with $A = A_c$ and $B =
B_c$.

We find in this way that eq.~\pref{teta} describes D$p$-brane
orbits for all $r \gg \ls$. For $g_s N \gg 1$ the parameters $A$
and $B$ are as given below eq.~\pref{teta}, and for $g_s N \lsim
1$ this becomes $A \approx A_c$ and $B \approx B_c$. Although one
might have expected the inclusion of the higher-order corrections
in powers of $v^2$ and $k/r^\td$ to more dramatically change the
form of the orbits, this does not happen.
Because the conditions $\eta = \omega = \zeta = 1$ which ensure
this simplicity for the brane/antibrane orbits ($q = - 1$) also
ensures the complete absence of an interaction potential between
branes ($q = + 1$), we conjecture that the independence of the
orbit shapes from receiving relativistic corrections arises only
for BPS branes and antibranes. A deeper understanding of when this
occurs, and whether there is a connection with the BPS nature of
the D$p$-branes involved, would clearly be worthwhile.

\subsubsection{Closed orbits}
We next demonstrate the existence of brane orbits which close on
themselves. Our interest here is not the circular closed orbits,
which close on themselves in a trivial way and which arise for
many kinds of branes. Rather we look here for the less trivial
situation where the orbits vary in both the angular and radial
directions, but these motions have commensurate periods.

The existence of such orbits is suggested by the following two
observations. First, we have seen that some relativistic branes
(like the BPS D$p$-branes in 10 dimensions) have precisely the
same orbits as has the nonrelativistic central-force problem.
Second, we know that motion in the nonrelativistic Coulomb
potential, $V \propto 1/r$, is along conic sections, with the
bound orbits being ellipses. We focus therefore on the case of
source D6-branes and a probe D6-antibrane, for which $\td = 7-p =
1$ and $q = -1$.

The probe-brane orbits are described by eq.~\pref{teta}, which for
$\td = 1$ may be integrated in closed form in terms of elementary
functions. When $A < 0$ the result is an ellipse
\begin{equation}
    r = \frac{a (1-e^2)}{1+ e \cos{\theta}} \, ,
\end{equation}
where the orbit's semimajor axis, $a$, and eccentricity, $e$, are
given by
\bea
    a &=& -\, \frac{B}{2A} = \frac{k \, \varepsilon}{2\left(2 -
    \varepsilon \right)}, \nonumber\\
    e &=& \left[ 1 + \frac{4A}{B^2} \right]^{1/2} =\phantom{B} \left[1 -
    \frac{4 \ell^2 \left(2 - \varepsilon \right)}{k^2
    \varepsilon^3} \right]^{1/2} \, .
\eea
For comparison, using the nonrelativistic values $A_c = 2
\varepsilon_c /\ell_c^2$ and $B_c = 2 k_c /\ell_c^2$ instead gives
the usual expressions $a_c = -k_c/(2\varepsilon_c)$ and $e_c =
\left[1 + 2 \, \varepsilon_c \ell_c^2/k_c^2 \right]^{1/2}$.

The time evolution along this orbit is obtained by integrating
equation (2.14), giving
\be
    t-t_0 =\ \int_{r_0}^r dx \; \frac{\Bigl[ \left(\varepsilon + q \right) x +
    \varepsilon \, k \, \Bigr]}{\sqrt{ \left( \varepsilon + 2 \, q \right)
    \varepsilon x^2 + \varepsilon^2 k \, x - \ell^2}}
\ee
which can be integrated to give
\be
    \Omega t = \psi - \tilde e  \sin\psi\ \label{tvspsi}
\ee
with the angle $\psi$ denoting the {\it eccentric anomaly},
defined by $r=a(1-e\cos\psi)$ or, equivalently,
\be \cos \psi =\ \frac{a-r}{ae} = \frac{k \varepsilon^2 -2 \,
\varepsilon (2-\varepsilon) \, r}{\sqrt{k^2 \, \varepsilon^4 - 4
\, \ell^2 \, \varepsilon (2 -\varepsilon)}} \, .\label{coseno} \ee
The angles $\psi$ and $\theta$ are explicitly related by $\tan
\frac{\theta}{2}=\sqrt{\frac{1+e}{1-e}}\tan{\frac{\psi}{2}}$.
Finally, the constant $\tilde{e}$ which appears in (\ref{tvspsi})
is:
\be \tilde e = \left( \frac{\varepsilon - 1}{3 - \varepsilon}
\right) \, e = \frac{\varepsilon - 1}{3 - \varepsilon} \, \sqrt{
1- \frac{4\ell^2 (2 -\varepsilon)}{k^2 \varepsilon^3}}\, . \ee
The period of the orbit is given by:
\be T\ =\ \frac{2\pi}{\Omega}\ =\ \pi k \left( \frac{3 -
\varepsilon}{2 - \varepsilon} \right) \sqrt{\frac{\varepsilon}{2 -
\varepsilon}}\ = \ {\pi}\ \sqrt{\frac{a}{2k}}\left( 2a + 3k\right)
\, . \label{period} \ee

It is instructive to compare this with the nonrelativistic Kepler
problem. In this case the time dependence is given by a relation
similar to eq.~\pref{tvspsi}:
\be
\Omega_c t  = \psi - e_c \, \sin{(\psi)}
\ee
where $\psi$ is related to $\theta$ as before. The orbital period
which results is $T_c = 2\pi/\Omega_c = 2\pi \sqrt{a_c^3/k_c}$.

We see that the D6-brane and the nonrelativistic problems share
the same functional form for both $t(\psi )$ and $r(\theta)$, but
the parameters which appear in the two solutions differ. (For
instance $\tilde e \neq e$ for the D6-brane, while $\tilde{e}_c =
e_c$ nonrelativistically.) This difference in parameters for the
two solutions implies a different relationship is obtained between
orbital features like $a$ and $e$ and the underlying physical
quantities, like $\varepsilon$ and $\ell$ and $m$.

Although the orbits have the same shape, the differences in their
parametric dependence on $\varepsilon$ and $\ell$ also implies
differences for how the period relates to the orbital shape and
size. For instance neither Kepler's second nor third laws apply
for D6-branes, since $T^2/a^3$ is not a universal quantity for all
probe brane orbits, and since conservation of $\ell$ does not
imply that the product $r^2 \, \dot{\theta}$ is a constant along
any one orbit.

\subsubsection{The Runge-Lenz vector}
Here we explore further the remarkable property that the closed
orbits for the nontrivial D6-brane system, described by the
lagrangian, (\ref{grangiano}), are identical to the orbits for the
standard nonrelativistic $1/r$ central force problem.

For the nonrelativistic Kepler problem, the fact that the orbits
are closed ellipses which do not precess is reflected by the
existence of an extra conserved quantity: the Laplace-Runge-Lenz
vector. Normally the existence of this vector is considered to be
a special consequence of the $1/r$ potential, which is not shared
by other central-force potentials.

Nonetheless, despite the fact that the D6-brane Lagrangian
contains many relativistic corrections as well as a potential
which is not purely proportional to $1/r$, its orbits are also
non-precessing ellipses. One might therefore wonder whether in
this case there is also a generalised Runge-Lenz vector whose
conservation ensures the absence of precession. We show in this
section, by explicit construction, that the answer to this
question is `yes'.

Since there are 3 spatial dimensions transverse to the D6 brane in
10 dimensions, it is convenient to phrase the analysis using the
angular momentum vector, which is dual to the conserved
antisymmetric tensor. Let us therefore consider the vector:
\be \label{RLAnsatz}
    {\bf \Lambda}\ \equiv \frac{d}{dt}\left(
    f(\varepsilon)\ {\bf p\wedge L}\right) \ = \ f(\varepsilon)\
    {\bf \dot p \wedge L}
\ee
where  ${\bf L = x\wedge p}$ is the conserved angular momentum
vector and for the moment $f(\varepsilon)$ is an unspecified
function of energy to be fixed below. The linear momentum is
defined as the canonical quantity
\be {\bf p}\ = \frac{1}{m} \frac{\partial L}{\partial{\bf \dot
x}}\ =\ \frac{1}{\sqrt{1-h \, v^2}} \ {\bf \dot x}, \ee
and
\bea {\bf \dot p}\ =  \ \frac{\partial L}{\partial {\bf x}}\ & = &
\ \frac{1}{h^2\sqrt{1 - h \, v^2}} \left[1 - \frac{h \, v^2}{2} -
q \sqrt{1- h \, v^2}\right]\ \frac{\partial h}{\partial r}\
\frac{{\bf x}}{r}\ \nonumber \\
& = &\ \frac{\varepsilon^2 \sqrt{1- h \, v^2}}{2m} \,
\frac{\partial h}{\partial r} \, \frac{{\bf x}}{r}, \eea
where in the last equality we used the expression for the energy
eq.~(\ref{energia3}) as well as  $q^2=1$. Using this expression to
evaluate the quantity ${\bf \dot p}$ in eq.~\pref{RLAnsatz} we
obtain:
\be {\bf \Lambda}\ = \ 2 f(\varepsilon)\, \varepsilon^2 \,
\frac{1}{r}\frac{\partial h}{\partial r}\ \left[ {\bf x } \wedge
\left({\bf x \wedge \dot x} \right) \right]. \ee
Now if we choose $f(\varepsilon)=1/\varepsilon^2$ and use the
identity
\be {\bf x} \wedge \left( {\bf  x \wedge \dot x} \right)\ =\ r \,
\dot r \, {\bf x} - r^2 \, {\bf \dot x}\ =\ -r^3 \, \frac{d}{dt}
\left(\frac{{\bf x}}{r}\right), \ee
we find
\be {\bf \Lambda}\ =\ \frac{1}{2} \frac{\partial h}{\partial r} \,
r^2 \,  \frac{d}{dt}\left(\frac{ {\bf x}}{r}\right) . \ee

Now comes the main point. Only for the particular case $h = 1 +
k/r$ is the right-hand-side of this last equation a total
derivative, and so can be used to make a conserved quantity ---
the generalised Runge-Lenz vector:
\be {\bf A}\ = \ \frac{1}{\varepsilon^2}\ {\bf p\wedge L}\ -\
\frac{k}{2}\ \frac{{\bf x}}{r}. \ee
Notice that, as expected, ${\bf A \cdot L}=0$, and
\be {\bf A\cdot x}\ =\ A \, r\cos\theta \ =\
\frac{\ell^2}{\varepsilon^2}\ -\ \frac{k \, r}{2} \ee
which exactly reproduces the elliptic trajectory $r(\theta)$ as
before, and sets $A=ke/2$.

As in the nonrelativistic case, the Runge-Lenz vector indicates a
further symmetry beyond rotation invariance, and because it lives
within the plane of the ellipse, pointing along the semimajor
axis, its conservation forbids the orbits from precessing.

\subsection{Quantum Dynamics}
We next describe the quantum behaviour which follows from the
lagrangian, \pref{grangiano}. We do so for its intrinsic interest,
even though all of our later brane applications will be well
within the classical domain.

Recall that for D$p$-branes the Hamiltonian is the following
function of the canonical momentum
\be {\cal H}\ = \ \frac{1}{h}\ \left(\sqrt{m^2+ h \, p^2} -
mq\right) \, .\ee
The Schr\"odinger equation for this Hamiltonian, ${\cal H}\Psi
=E\Psi$, therefore becomes
\be \frac{1}{h}\ \left(\sqrt{m^2 + h \, p^2} - m \, q\right) \Psi
\ = \ E \Psi \, . \ee
This may be written in a more useful form by isolating and
squaring the square root on the left-hand-side, leading to
\be \left(m^2 + h \, p^2\right)\ \Psi\ =\ \left(E h + m q
\right)^2\ \Psi \, .\ee
Because we have squared ${\cal H}$ in what follows we will find
solutions having both signs for $E$. Since ${\cal H}$ is
manifestly positive for $q = \pm 1$, we must drop all solutions
for which $E < 0$ (notice that the time-dependent Schr\"odinger
equation will include two time derivatives indicating positive and
negative frequencies).

For $q^2 = 1$ and $h = 1+ k/r^\td$ this can be further manipulated
to be written as:
\be \label{SchEq}
    \left(p^2- \frac{E^2 \, k}{r^\td}\right)\ \Psi = E (2mq + E)
    \Psi \, .
\ee
The utility of this last form comes from its similarity with the
corresponding nonrelativistic central-force problem:
\be
    \left(\frac{p^2}{2\mu}- \frac{g^2}{r^\td} \right)\ \Psi\ = \
    E_c \Psi \, ,
\ee
which becomes identical to eq.~\pref{SchEq} after the replacements
$\mu \to 1/2, g^2 \to E^2 k$, and $E_c \to E (E + 2qm)$.

In particular, for D6 branes we have $\td = 1$, and so
eq.~\pref{SchEq} has the same form as the Schr\"odinger equation
for the Hydrogen atom, for which the solutions have been known
since antiquity. In this case the energy eigenvalues consist of a
continuum, $E_c \ge 0$, plus a discrete spectrum of bound states
labelled by the quantum number $n = 1,2,...$: $E_c = - g^4 \,
\mu/(2 n^2)$.\footnote{Recall our use of fundamental units, for
which $\hbar = c = 1$.}

\FIGURE{
\centering
\epsfxsize=3in \hspace*{0in}\vspace*{.2in}
\epsffile{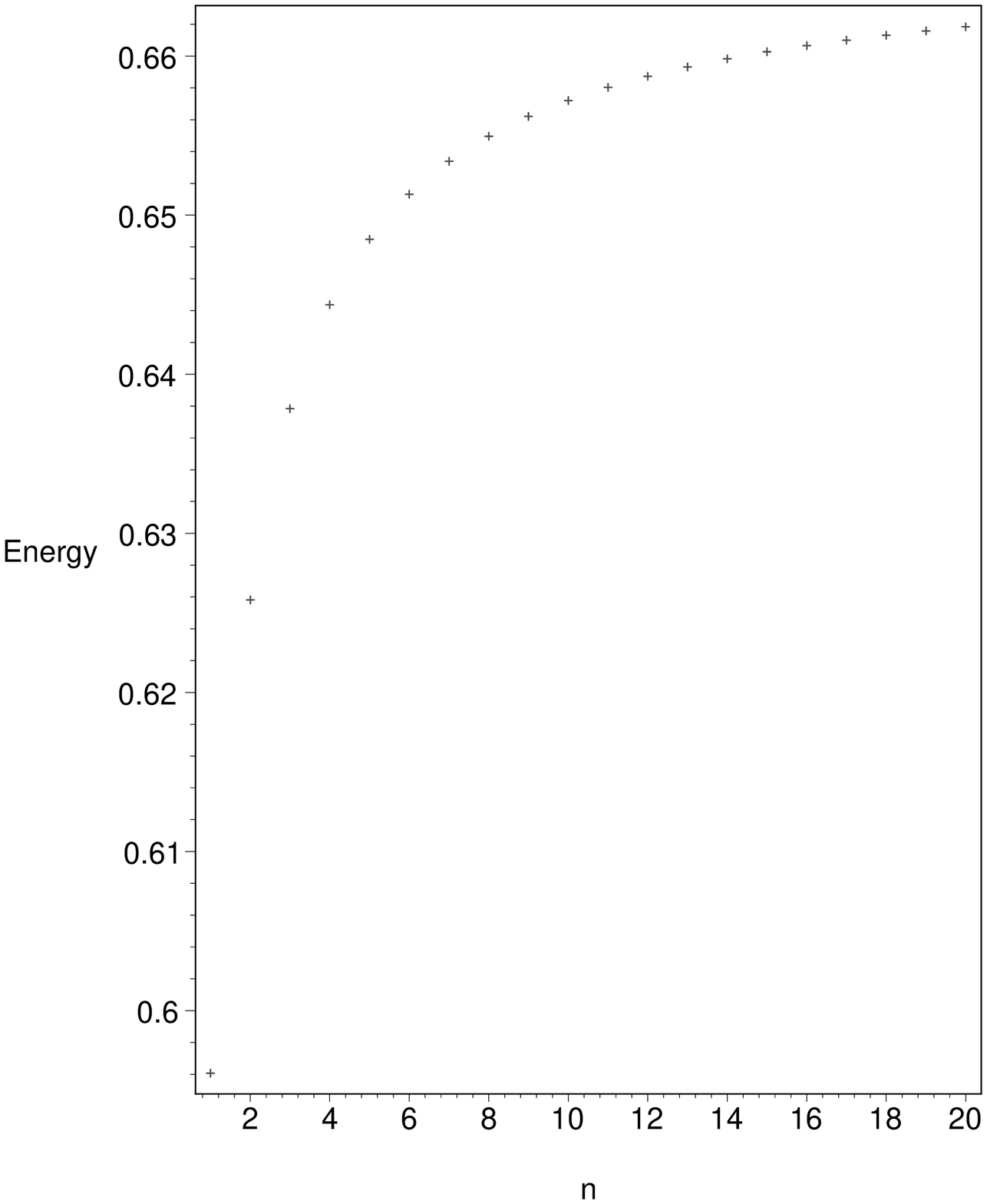}
\caption{\small
The energy of the first 20 bound states of the branonium.
We have taken $m= 1/3$ and $k^2 = 4/3$.}
\label{spec}
}

Making the substitutions which convert this spectrum to the
D6-brane problem, we find the following energy spectrum:
\bea
    E(E + 2qm) &\ge& 0 \, , \qquad\qquad \hbox{(continuum)} \\
    E(E+2qm) &=& - \frac{E^4 k^2}{4n^2} \, , \qquad\qquad
    \hbox{(discrete)}. \nonumber
\eea
Requiring an energy spectrum which is bounded from below implies
the continuum branch becomes $E \ge 0$ if $q = 1$ and $E \ge 2m$
if $q = -1$.

The energies for the discrete spectrum can be explicitly solved to
give the following expression in terms of the level $n$ and the
constant $k$:
\be \label{DiscSpec}
    E(n) = \left( \frac{8n^2}{k^2} \right)^{1/3} \left[ \left(
    \sqrt{\frac{m^2}{4} + \frac{n^2}{27 k^2}} - \, \frac{qm}{2}
    \right)^{1/3} - \left( \sqrt{\frac{m^2}{4} + \frac{n^2}{27 k^2}} +
    \frac{qm}{2} \right)^{1/3} \right] \,.
\ee
Notice that the square bracket in this expression is negative if
$q = +1$ and is positive if $q = -1$. Consequently for $q = +1$
these discrete eigenvalues are spurious and must be discarded. For
$q = -1$ the discrete eigenvalues are positive, and approach the
continuum ($E = 2m$) as $n \to \infty$. The first 20 energy levels
which follow from eq.~\pref{DiscSpec} are plotted against $n$ in
Figure \ref{spec}.

The characteristic size of the corresponding discrete energy
eigenstates in the transverse dimensions is set by the D6-brane
analog of the Bohr radius: $a_B = 1/(g^2 \, \mu)$. After the
appropriate substitutions this becomes
\be a_B\ = \frac{2}{E^2 \, k} \, . \label{Bohr} \ee

For subsequent applications it is interesting to see how these
expressions behave in the decompactification limit, for which the
spatial world-volume, $V_6$, of the brane becomes very large.
Because $m = T_6 V_6$, this corresponds to the large-$m$ limit,
with $n$ and the interaction constant $k = c_6 g_s N \ls$ held
fixed. We see that in this limit the energy and the Bohr radius of
the bound states scales like
\be
    E \propto \frac{m^{1/3}}{k^{2/3}} \, , \qquad a_B \propto
    \frac{k^{1/3}}{m^{2/3}} \, .
\ee
We see from this that the quantum bound states are not reliable in
the limit of large $V_6$, for fixed $N$, because in this limit
they imply $a_B \ll \ls$.

\subsection{Compactification}
We have found potentially long-lived orbits for probe antibranes
moving in the field of $N$ source branes, provided $\td = 1$ and
so they move in 3 transverse dimensions. For 10 dimensions
overall, this is what singles out D6-branes for special attention.
For later applications to brane-world cosmology we will be
interested in branes whose world-volume is 4 dimensional, and so
we pause here to consider how this may be achieved from the above
solutions by compactification.

\subsubsection{Toroidal Reduction}
Our goal in this section is to dimensionally reduce the action,
eq.~\pref{eq:EFbulkaction}, from $D$ to $D' = D-1$ dimensions. The
point is to present a consistent lower-dimensional truncation of
this action, in the sense that any solution of the
lower-dimensional equations of motion are also solutions of the
equations of the full system, eq.~\pref{eq:EFbulkaction},
following the construction of ref.~\cite{Kelly}.

To this end imagine that one of the coordinate directions, denoted
$z$, is periodic. In this case we write the $D$-dimensional
metric, $ds^2_D$, in terms of a $D'$-dimensional metric,
$ds^2_{D'}$ plus a scalar field, $\varphi$, and a vector field,
${\cal B}_M$ as follows:
\be \label{KKansatz1}
    ds^2_D = e^{2 \hat{a} \varphi} \, ds^2_{D'} + e^{2 \hat{b}
    \varphi} \, (dz + {\cal B}_M dx^M)^2 \, ,
\ee
with $\hat{b} = -(D'-2)\, \hat{a}$ and $\hat{a}^2 =
1/[2(D'-1)(D'-2)]$ chosen to ensure canonical Einstein and
$\varphi$ kinetic terms. The $D$-dimensional $(n-1)$-form field,
$A_{[n-1]}$, is similarly decomposed into $D'$-dimensional
$(n-1)$- and $(n-2)$-form fields, $B_{[n-1]}$ and $B_{[n-2]}$,
according to
\be \label{KKansatz2}
    A_{[n-1]} = B_{[n-1]} + B_{[n-2]} \wedge dz \, .
\ee

With these choices the action, eq.~\pref{eq:EFbulkaction},
dimensionally reduces to
\bea
    S_{D'} &=& - \int d^{D'}x \, \sqrt{-g_{D'}} \, \left[ \sfrac12 \, R_{D'}
    + \sfrac12 \, (\partial \phi)^2 + \sfrac12 \, (\partial
    \varphi)^2 + \sfrac14 \, e^{-2(D'-1) \hat{a} \varphi} \, {\cal
    B}_{[2]}^2 \right. \nonumber \\
    && \qquad \left. + \frac{1}{2 n!} \, e^{-2(n-1)\hat{a} \varphi
    + \alpha \, \phi} G^{'2}_{[n]} - \frac{1}{2(n-1)!} \,
    e^{2(D'-n) \hat{a} \varphi + \alpha \, \phi} \, G^2_{[n-1]}
    \right] \, ,
\eea
where the field strengths appearing in this expression are defined
by
\be
    {\cal B}_{[2]} = d{\cal B}, \qquad G_{[n-1]} = dB_{[n-2]},
    \qquad G'_{[n]} = dB_{[n-1]} - G_{[n-1]} \wedge {\cal B} \, .
\ee

This lower-dimensional action may be consistently truncated in
three distinct ways, in the above sense of consistency. The three
ways are obtained by setting to zero all fields except the
$D'$-dimensional metric and one of the following pairs of
form-fields and canonically-normalized scalars \cite{Kelly}:
\bea    \label{lowerdim}
    \hbox{Option A:} \qquad &&{\cal B} \quad \hbox{and} \quad
    \phi_1' = \varphi \nonumber \\
    \hbox{Option B:} \qquad &&B_{[n-2]} \quad \hbox{and} \quad
    \phi'_{n-2} = \frac{2(D'-n) \, \hat{a}
    \varphi + \alpha \phi}{\sqrt{4(D'-n)^2 \, \hat{a}^2 + \alpha^2}} \\
    \hbox{Option C:} \qquad &&B_{[n-1]} \quad \hbox{and} \quad
    \phi'_{n-1} = \frac{-2(n-1) \, \hat{a} +
    \alpha \phi}{\sqrt{4(n-1)^2 \, \hat{a}^2 + \alpha^2}} \, .\nonumber
\eea

Each of these reductions produces a $D'$-dimensional action which
again has the form of eq.~\pref{eq:EFbulkaction}, but with
parameter $\alpha = \alpha_{D'}$ now given in terms of the
higher-dimensional parameter $\alpha_D$ by:
\bea
    \hbox{Option A:} \qquad &&\alpha_{D'} = -2(D'-1) \alpha_D \qquad
    \hbox{implying} \qquad \Delta' = 4 \, , \nonumber\\
    \hbox{Option B:} \qquad &&\alpha_{D'} = \sqrt{4(D'-n)^2 \hat{a}^2 + \alpha_D^2}
    \qquad
    \hbox{implying} \qquad \Delta' = \Delta \, , \\
    \hbox{Option C:} \qquad && \alpha_{D'} = \sqrt{4(n-1)^2 \hat{a}^2 + \alpha_D^2}
    \qquad
    \hbox{implying} \qquad \Delta' = \Delta \, , \nonumber \, . \eea

The two main choices are now whether to apply this reduction to
the case where the compactified dimensions are transverse or
parallel to the source and probe branes' world volumes, and so we
now consider both of these options.

\subsubsection{Parallel Dimensions}
The simplest option is to compactify dimensions which are parallel
to the branes' world volumes by wrapping the branes about a torus.
Wrapping D6-branes about a 3-torus of radius $R$ in this way would
produce a brane world which is effectively 4 dimensional when
viewed on scales large compared with $R$ by virtue of the standard
Kaluza-Klein mechanism. Such a toroidal compactification is
straightforward since the solutions we have considered are
invariant under translations along the directions parallel to the
branes. In particular, it is clear that the solution,
eq.~\pref{Dbrane}, remains a solution to the full
higher-dimensional field equations if some of the coordinates
$x^\mu$ are required to be periodic.

Writing $x^p = z$ and directly comparing eqs.~\pref{Dbrane} with
the Kaluza-Klein ans\"atze, eqs.~\pref{KKansatz1} and
\pref{KKansatz2}, allows the field configurations to be written
purely in terms of lower-dimensional fields as follows:
\be \label{Dbrane'}
    ds^2_{D'} = h^{-\tgamma'} \, dx^2_{p'} + h^{\gamma'} \, dy^2
    \, ,\quad e^{\phi} = h^{\kappa} \, ,\quad e^{\varphi} = h^\rho \, ,
    \quad (B_{[n-2]})_{01...p-1} = \zeta (1 -  h^{-1}) \, ,
\ee
where $\kappa$ and $\zeta$ are as before ({\it c.f.}
eq.~\pref{braneconsts}) and
\be
    \rho = \frac{\tgamma}{2(D'-2) \, \hat{a}}, \quad
    \gamma' = \gamma - 2 \, \hat{a} \, \rho, \quad \tgamma' =
    \tgamma + 2 \, \hat{a} \, \rho .
\ee
Notice that although this solution resembles truncation option B
of eq.~\pref{lowerdim}, it is {\it not} simply a special case of
the solutions we have considered so far, with the replacements $D
\to D'$, $p \to p'$ and $\alpha \to \alpha'$. It is not because in
this solution {\it both} of the scalar fields $\phi$ and $\varphi$
vary nontrivially in the transverse dimensions. (In particular,
the combination orthogonal to $\phi'_{n-2}$ of eq.~\pref{lowerdim}
does not vanish.)

By starting from a given $p$-brane solution in $D$ dimensions and
successively compactifying $k$ of the parallel dimensions we
obtain new configurations which describe $p-k$ branes moving in
$D-k$ dimensions, without changing the number of transverse
dimensions and so also not changing $\td$. (These are the
`diagonal' reductions of ref.~\cite{Kelly}.) By construction this
reduction leaves the values of the parameters $\eta$, $\omega$ or
$\zeta$ which appear in the probe brane action unchanged.

Reducing the BPS D$p$-branes in 10 dimensions in this way then
leads to a family of $p-k$ branes moving in $10 - k$ dimensions,
for which $\eta = \omega = \zeta = 1$, and so for which the
orbital properties are particularly simple. The choices $p=6$ and
$k=3$ give the particularly interesting case of 3-branes moving in
3 transverse dimensions. For these $\td = 1$ and so the orbits are
the closed ellipses described earlier.

\subsubsection{Transverse Dimensions}
To this point we have treated the transverse dimensions as if they
were infinite in extent. Although this type of geometry might have
brane-world applications if gravitons can be confined to our brane
(perhaps through the warping of spacetime), we need not rely on
this when exploring the various implications of branonium. If not,
we must also address the dimensional reduction of the transverse
dimensions.

Reduction of the transverse dimensions is trickier because the
field configurations in these directions are not translationally
invariant. For BPS branes, such as the 10-dimensional D$p$-branes,
this can be dealt with using a generalization of the solutions we
have considered to source branes which are not all located at the
same point. This more general solution is obtained using the same
expressions, eqs.~\pref{Dbrane}, as before but with the harmonic
function of eq.~\pref{harmonic} replaced by the new expression
\be
    h(y) = 1 + \sum_i \frac{k}{|{\bf y} - {\bf y}_i |^\td} \, ,
\ee
where $|{\bf y} - {\bf y}_i|^2 = \delta_{mn} (y-y_i)^m (y -
y_i)^n$. Given this solution a toroidal compactification of the
transverse dimensions may be performed simply by including the
appropriate image charges for the source branes, which serve to
ensure that the transverse dimensions become periodic
\cite{Kelly}.

There are two cases of this kind of reduction which concern us,
depending on the relative size of the compact dimensions and the
size of the brane orbits which are of interest. If some of the
transverse dimensions are very small compared to the orbital sizes
of interest, then we may integrate out these directions and work
in the limit where their radius is taken to zero. If $k$
transverse directions are so reduced it has the effect of reducing
$D \to D' = D - k$ without changing $d$, and so gives new brane
solutions for which $\td \to \td' = \td - k$. (These are the
`vertical' reductions of ref.~\cite{Kelly}.)

For the brane orbits of previous sections we must take $\td > 2$,
however, and so at least some of the transverse dimensions must
have radii, $R$, which are larger than the orbital sizes, $a$, of
interest. In this case the orbital analysis of the previous
sections must be considered to be only approximately valid, up to
corrections which are of order $(a/R)^\td$ for $a \ll R$. For $a
\sim R$ the influence of the image sources can significantly alter
the probe-brane motion. In our later applications we imagine at
least three of the transverse dimensions to be stabilized with
sufficiently large radii.

\FIGURE{ \centering \epsfxsize=3in \hspace*{0in}\vspace*{.2in}
\epsffile{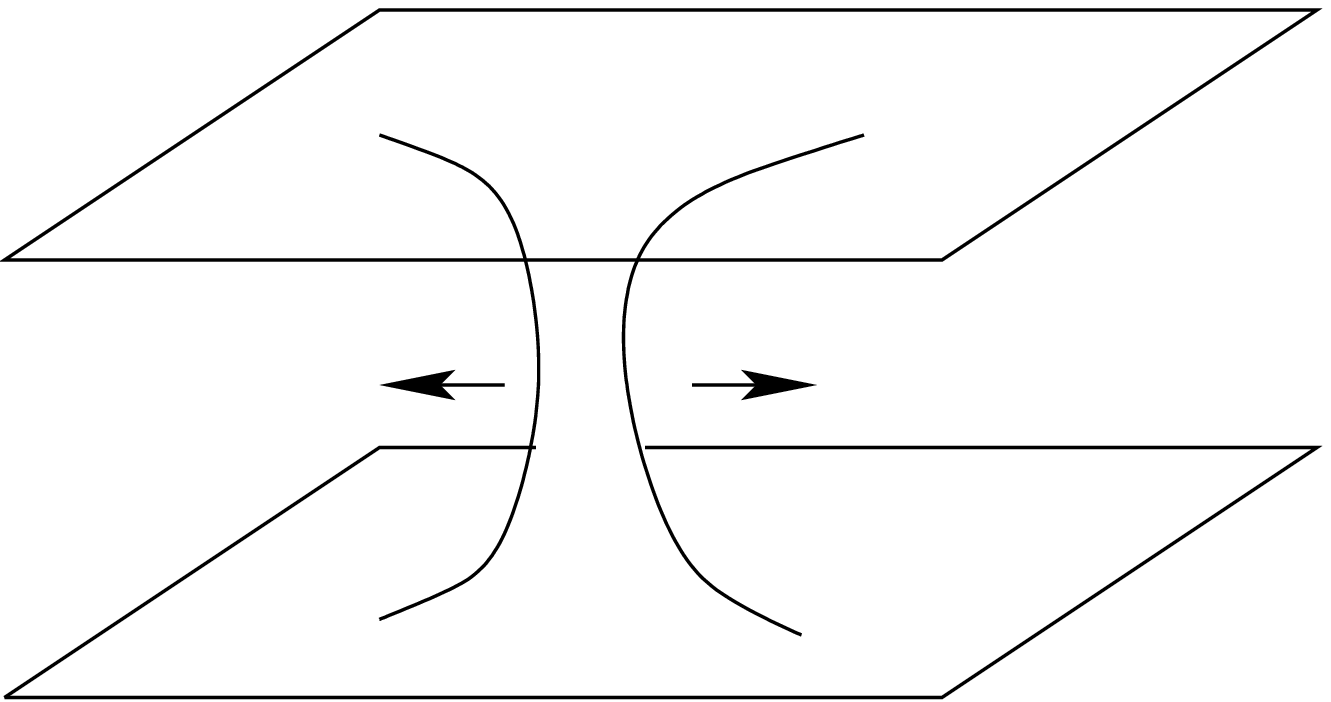} \caption{\small Nucleation of a throat
between a brane and and antibrane.}}\label{throat}

\section{Stability}
We now turn to the stability of the orbits described in the
previous section (see \cite{stab} for a related discussion). There
are several kinds of potential instability to consider. To this
point we have treated the probe-brane motion as if the branes were
perfectly rigid, with only their centres of mass free to move. The
first instability issue to address is the extent to which this is
a good approximation, inasmuch as other kinds of brane motion
might become excited by the movement of the centre of mass. To
address this at the classical level we look for tachyonic modes
among the linearized fluctuations of the brane position,
$y(\xi,t)$, about the field configurations of interest. We argue
that this kind of instability need not be present so long as the
source and probe branes are kept at distances which are much
larger then the string length.

The second stability issue asks whether the brane motion produces
radiation of massless bulk and/or brane fields, thereby radiating
away energy and so causing the brane orbits to decay. This kind of
instability does occur, and we here estimate how long it takes to
destroy a given brane orbit.

All told we identify three separate kinds of instabilities to
which orbiting branes are subject.
\begin{itemize}
\item Brane annihilation, due to the perturbative open string tachyon.
If the brane and antibrane get within a string length or so of one
another, then they can annihilate with the antibrane being
absorbed by the $N$ source branes, ultimately leaving a
configuration of $N-1$ branes. The energy released is then
radiated into modes in the bulk and on the remaining source
branes. We do not pursue this in detail because our present
interest is in branes which are sufficiently well separated as to
avoid this. Notice that, depending on their number and size, for
compact transverse dimensions sufficient separation of the branes
can require them not to be too differently wound around the extra
dimensions.
\item A non-perturbative brane decay is possible, \`a la Callan-Maldacena
\cite{cm97}. These are solutions to the Euclidean equations of
motion. They can be interpreted as sphalerons associated to the
decay of a brane-antibrane system. A throat is nucleated that
joins the brane-antibrane pair. Then it expands and transforms the
brane world-volume into kinetic energy of the expansion front (see
figure \ref{throat}). We do not pursue this instability further
since its nucleation probability is proportional to $e^{-1/g_s}$,
and so is exponentially small for small string coupling.
\item Radiation into bulk modes, such as the graviton
and dilaton, or brane modes can occur and arises at low orders in
the string coupling. We expect this to be the dominant instability
for large interbrane separations and weak couplings, and estimate
the time for orbital decay below.
\end{itemize}

We will then establish first the stability of the system against bending
and concentrate later only on the instability due to radiation.

\subsection{Stability Against Brane Bending}
%
\FIGURE{ \centering \epsfxsize=4in \hspace*{0in}\vspace*{.2in}
\epsffile{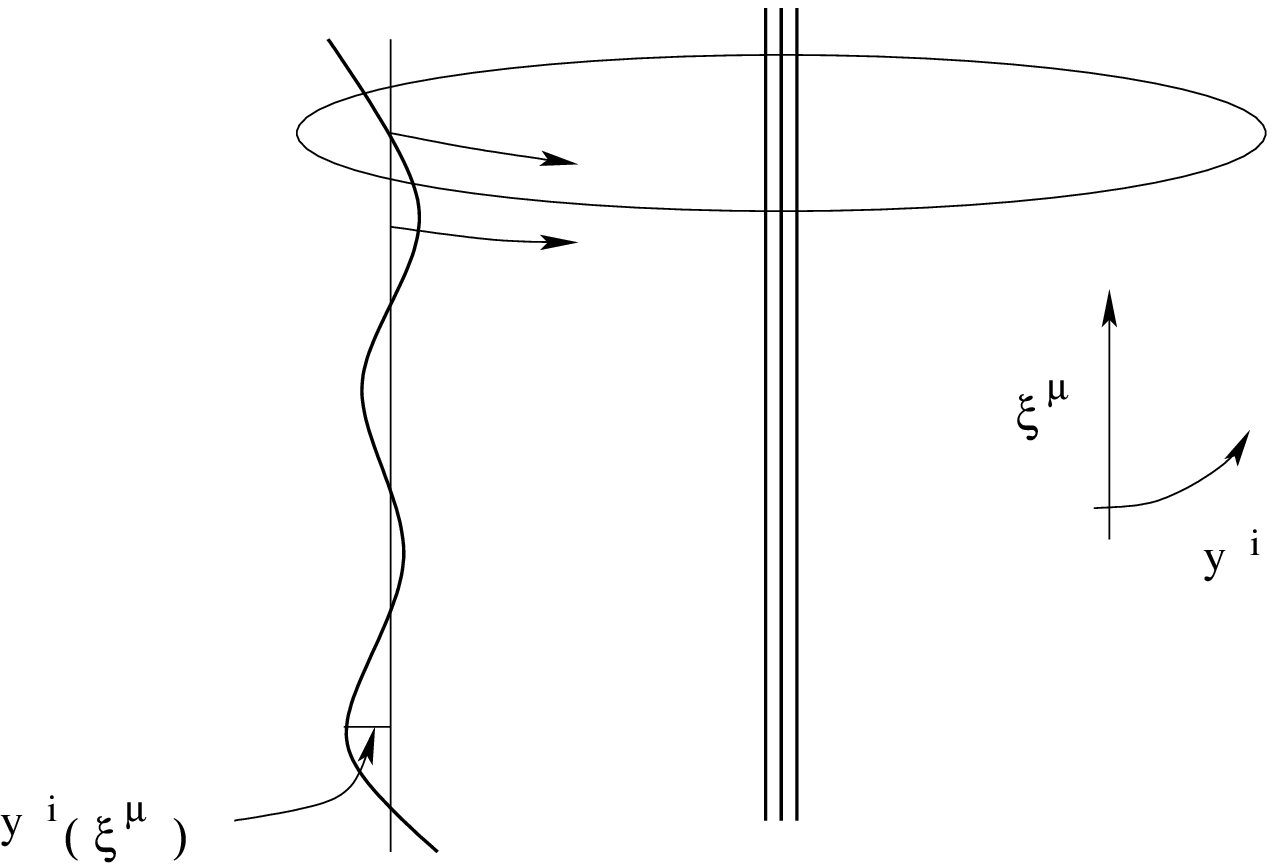} \caption{\small The fields $y^i(\xi)$ on
the brane parameterise its fluctuations. Instabilities
corresponding to brane bending or the classical orbits can be
analyzed from the mass of these modes in the quadratic
approximation.} \label{bending} }
%
Instead of considering only explicitly straight and parallel
branes, imagine allowing them to bend so that the probe-brane
position relative to the centre-of mass is $y^i(t,\xi^{\mu}) = y^i
- y^i_0$, where as before $\{ t,\xi^{\mu} \}$ are the coordinates
along the brane (see figure \ref{bending}). Nontrivial dependence
of $y^i$ on $\xi^\mu$ corresponds to brane bending, while
perturbations depending only on time, i.e. $y^i := y^i(t)$,
describe the rigid motion of the brane centre of mass.

A stability analysis is performed by expanding the action about a
particular background solution, $y^i = y^i_0(t) + \delta
y^i(\xi,t)$, with the background $y_0(t)$ taken to describe the
rigid brane motion described in the previous sections. The
expansion of $L$ about this solution defines the system's {\it
Routhian} \cite{g}, which to quadratic order is
\begin{equation}
    L_{eff} = - R = \frac{1}{2} A^{\mu \nu}_{ij} \;
    \partial_{\mu}{\delta y}^i \, \partial_{\nu}{\delta y}^j
    + B^{\mu}_{ij}\; {\delta y}^i \,
    \partial_\mu \delta y^j + \frac{1}{2}C_{ij}\;
    \delta y^i \delta y^j \, ,
\end{equation}
where
\be
    C_{ij}[y_0(t)] = \left. \frac{\partial^2{R}}{\partial{y^i} \,
    \partial{y^j}} \right|_{y^{0}}, \qquad \; B^{\mu}_{ij}[y_0(t)] =
    \left. \frac{\partial^2{R}}{\partial(\partial_{\mu}{y}^i) \,
    \partial{y^j}} \right|_{y^{0}} \,
\ee
and
\be
    A^{\mu \nu}_{ij} [y_0(t)] =
    \left. \frac{\partial^2{R}}{\partial(
    \partial_{\mu}{y}^i) \, \partial(\partial_{\nu}{y}^j)}
    \right|_{y^{0}} \, .
\ee
One must analyse this effective Lagrangian looking for unstable
perturbations, given the known functions for the quantities
$A_{ij}^{\mu\nu}$, $B^\mu_{ij}$ and $C_{ij}$.

For the present instance, we may for simplicity focus on the
large-separation limit and so take the non-relativistic
approximation
\begin{equation} \label{ApproxL}
    L \approx m \left[ \frac{v^2}{2} + (\eta - \hat{q})
    \frac{k}{r^{\tilde{d}}} \right] \, .
\end{equation}
The stability analysis greatly simplifies in this case because
simplicity of the derivative terms implies the quantities
$A_{ij}^{\mu\nu}$ and $B^\mu_{ij}$ are independent of fields, and
so are also independent of time. As a result it also follows that
$B^\mu_{ij} = 0$.

The simplicity of the kinetic terms allows a further
simplification, since it implies gradients contribute a strictly
positive amount to the energy density. This means that it is the
zero-momentum modes in a Fourier decomposition, $\delta y^i(k)
\propto e^{i {\vec k} \cdot {\vec \xi}}$, which are the most
unstable. It follows that if these zero-momentum modes are stable,
then so must be all of the modes having nonzero momentum. This
stability of nonzero momentum modes has a simple physical
interpretation. On one hand, if the probe brane were to bend
slightly towards the source branes, then tidal forces from the
source branes will act to increase the amplitude of this bending.
The amount of this tidal disturbance will fall with increasing
brane separation. On the other hand, the brane tension resists all
such bending and easily wins so long as the interbrane separation
is large compared with the string length.

We now specialize to the stability analysis for the zero-momentum
modes in a purely time-dependent background. The effective
Lagrangian for the perturbations $\delta y^i = y^i - y^i_0$ then
simplifies to
\begin{equation}
    L_{eff} = - R = \frac{1}{2} {\cal A}_{ij}
     {\delta \dot y}^i {\delta \dot y}^j +
    {\cal B}_{ij} {\delta \dot y}^i \delta y^j +
    \frac{1}{2} {\cal }C_{ij}  \delta y^i \delta y^j \, ,
    \end{equation}
where ${\cal C}_{ij} = \left. \frac{\partial^2{R}}{\partial{y^i}\,
\partial{y^j}} \right|_{y_{0}}$, ${\cal B}_{ij} = \left.
\frac{ \partial^2{R}}{\partial{\dot{y}^i} \,
\partial{y^j}} \right|_{y_{0}}$ and ${\cal A}_{ij} = \left.
\frac{\partial^2{R}}{\partial{\dot{y}^i}\partial{\dot{y}^j}}
\right|_{y_{0}}$.
Applied to the lagrangian, eq.~\pref{ApproxL}, we get the
following Routhian expansion for the perturbations:
\begin{equation}
\frac{L_{eff}}{m} = - R  \approx \frac{1}{2} \Bigl[ (\delta
\dot{r})^2 + r_0^2 \, (\delta \dot{\theta})^2 + (\delta
\dot{y}^i)^2 \Bigr] + \frac{1}{2} \left[ (\eta - \hat{q}) \frac{k
\tilde{d} (\tilde{d} +1)}{r_0^{\tilde{d}+2}} -  3
\left(\frac{\ell}{r_0^2} \right)^2 \right] (\delta r)^2 \, .
\end{equation}

Only the radial mode appears nontrivially in these manipulations,
leading to the `mass':
\begin{equation}
M_{r}^2 = - m (\eta - \hat{q})\frac{k \tilde{d} (\tilde{d}
+1)}{r^{\tilde{d}+2}} + \frac{3\ell^2}{r^4} \, .
\end{equation}
Our interest is in $\eta - \hat{q} > 0$, such as occurs for probe
D$p$-antibranes, and so the first term in $M_r^2$ is negative
while the second term is positive. Which term dominates at large
$r$ depends on the relative size of the power of $r$ in the two
terms, and so $M_r^2 < 0$ for large $r$ when $\tilde{d} < 2$ and
$M_r^2 > 0$ for large $r$ when $\td > 2$. In the intermediate case
$\tilde{d} = 2$ the sign of $M_r^2$ at large $r$ is the same as
the sign of the combination $3 \ell^2 - mk(\eta - \hat{q}) \,
\td(\td + 1)$. For the D$p$-branes this means that we have
classical instability at large $r$ for $p >5$ and classical
stability for $p = 6$. This is as expected since we know that
deviations from the classical orbit in the Kepler problem for $\td
= 1$ are purely oscillatory.

Since D6-branes can be supersymmetric, the oscillations about the
stable configuration fill out an entire supermultiplet of the
appropriate unbroken supersymmetries. Seen from 4 dimensions on
the brane, this could be an $N=4$ SYM multiplet, depending on how
many supersymmetries are broken by the compactification.

A similar analysis can be done to study the stability of the
brane/antibrane system against developing a relative rotation as
their centres of mass move. This does not occur because parallel
branes furnish a local minimum of the potential energy as a
function of angle \cite{JoesBigBookoString}. For branes embedded
in infinite 10D spacetime, the oscillation time scale for small
rotational perturbations can be quite large, due to the necessity
to rotate an extremely long brane. In this case one should check
how this oscillation time scale compares with the orbital period
calculated above. For real applications this time scale is less
important, however, because the branes then move within compact
transverse dimensions. For toroidal compactifications, for
instance, branes cannot rigidly rotate relative to one another
without a large energy cost. This cost arises because local
rotation of the branes necessarily requires the branes to bend
somewhere, due to the necessity that they cannot change their
windings about the cycles of the torus.

We do not expect this result to change when relativistic
corrections are taking into account, and so we expect the
relativistic system to also be stable (for $\td = 1$) provided the
interbrane separation is kept well away from the string length. We
now turn to the orbital energy loss due to radiation of various
string modes.

\subsection{Radiation into the Bulk}
The most important decay mode for large orbits at weak coupling is
energy loss through radiation into the various massless fields in
the problem. For the classical motion which we are considering,
this radiation causes the orbits to slowly shrink in size, much as
the binary-pulsar orbit has been observed to do due to
gravitational radiation. Once the orbital size reaches the string
scale the much quicker open-string-tachyon instability takes over,
causing a catastrophic annihilation of the antibrane with the
brane. Our purpose in this section is to estimate the time which
is required before this catastrophe takes place.

\subsubsection{Radiation Rates}
Unlike the situation for the binary pulsar, the moving probe
antibrane has several types of charges which couple its motion to
fields whose radiation could carry away energy, including at the
very least the bulk gravitational, $(n-1)$-form and dilatonic
fields. For nonrelativistic motion, when all other things are
equal, the main carriers of radiated energy are the lower-spin
fields \cite{ac99}.\footnote{This observation also applies to the
binary pulsar in the presence of very light scalar and vector
particles, as may be seen, for example, in ref.~\cite{Oldie}.}
This is because in this limit spin-0, -1 and -2 particles
respectively dominantly couple to the orbiting brane's monopole-,
dipole- and quadrupole-moments, with each higher multipole
suppressed by additional small factors of powers of $1/r$. We
therefore base our estimate on radiation into a bulk scalar field
like $\phi$.

Suppose there are $D$ dimensions which are large compared with the
brane orbits, of which $d$ are parallel to the various branes. The
case of special interest is $d=4$ and $\td = 1$, which corresponds
to $D = 7$, but for the present purposes we need not restrict
ourselves to this case. We imagine the probe brane orbit to be
circular, with orbital radius $r$ and orbital frequency $\Omega$.

Recall that $\phi$ appears in the brane action through the overall
factor $e^{-\phi} \, m$, with (as usual) $m = T_p \, V_p$, where
the spatial volume of the brane enters because of the overall
translation invariance within the brane dimensions. This action is
given in units where the scalar and Einstein kinetic terms are $S
= -\, \sfrac12 \, \int d^Dx \, \sqrt{-g} \, \Bigl[R + (\partial
\phi)^2 \Bigr]$, and so we see that the $\phi$-brane coupling is
proportional to $\kappa_T \, m$, where $\kappa_T^2 = 8 \pi
G_D/V_p$ relates $\kappa_T$ to the $D$-dimensional Newton's
constant. Furthermore, since for slowly-moving objects
Lorentz-invariance precludes any radiation unless the source
accelerates, the $\phi$-emission amplitude should be proportional
to the brane acceleration, $a_b$.

Keeping track of dimensions and factors of $2\pi$, the power
radiated into scalars per unit brane volume therefore has the
schematic form
\be \label{Pexp}
    P \sim 2 \pi  \, \kappa_T^2 \, m^2 \, a_b^2
    \int \delta(E - \Omega) \frac{d^{D-d}k}{(2 \pi)^{D-d} \, E^2} \, ,
\ee
where the final integral is over the phase space of the emitted
scalar. The Dirac $\delta$-function is meant to express the fact
that energy conservation demands the dominant energy of the
outgoing scalars to be set by the dominant frequency, $\Omega$, of
the time-dependent source. It is this energy which sets the scale
of the phase space integrations. The power of $E$ in the integral
is chosen on dimensional grounds.

We estimate the phase-space integration by replacing it with the
appropriate power of the relevant energy scale, $\Omega$, and by
counting the factors of $2\pi$ which are produced by the angular
integrations in the transverse dimensions. This leads to the
estimate that the phase-space integral is of order $(2
\pi)^{[(D-d)/2]} \, \Omega^{D-d-3}$, where $[k/2]$ denotes the
integer part of $k/2$. We find in this way the following estimate
for the total power radiated:
\be
    P \sim \kappa_T^2 \, m^2 \, a_b^2 \, \left(
    \frac{\Omega^{D-d-3}}{(2 \pi)^{{\cal P}}} \right)\, , \ee
with ${\cal P} = D-d -1- \left[\sfrac12 (D-d) \right]$.

For three transverse dimensions we would have $D-d=3$ and so
$[\sfrac12(D-d)] = 1$. This would be appropriate for point
particles (0-branes) moving in 4 dimensions, in which case we
would have $D=4$ and $d = 1$ and $\kappa_T = \kappa_4$ is the
ordinary 4D gravitational coupling. The above estimate then
reproduces the results of more careful calculations, which give:
\be P = \frac{1}{8 \pi}\, (\kappa_4 m)^2 a_b^2  \, .\ee
This same formula applies equally well to the case of more general
branes orbiting in 3 transverse dimensions, provided we use the
appropriate choice for $\kappa_T$
\be \label{BulkRad}
    P = \frac{1}{8\pi} \, \kappa_T^2 \, m^2 \, a_b^2 \,=
    \frac{1}{8\pi} \, \kappa_D^2 \, T_p^2 \, a_b^2 \, V_p \, .
\ee

Notice that in all cases $P$ scales with the spatial volume of the
orbiting branes, as is required by their translational symmetry,
and so the power radiated per unit energy is independent of $V_p$.
As a consequence of this $V_p$ drops out of the time required for
an orbit to decay, as we shall shortly see in more detail. Also of
note is the dramatic phase-space suppression (by powers of $\Omega
\, \ls$) which arises when there are more than 3 transverse
dimensions.

\subsubsection{Orbital Decay Time}
We now turn to the time required for a circular orbit of radius
$r$ to decay, given the above estimate for the power radiated. We
focus on the case $\td = 1$, since it is for this case that these
stable orbits are most likely to exist. For the present purposes
we write the power radiated as $P = K \, m^2 \, a_b^2$, with $K$
containing the various dimension-dependent constants.

To evaluate the acceleration it suffices to work in the limit of
large orbits and small velocities, where the brane Lagrangian
becomes (up to an additive constant)
\be L = \frac{mv^2}{2} - V(r) \ee
where $V = - 2k \, m/r$. The acceleration is then
\be a_b^2 = \frac{1}{m^2} \left( \frac{\partial V}{\partial r}
\right)^2 = \frac{4  k^2}{r^4} \, , \ee
and so the power radiated becomes
\be P = - \, \frac{dE}{dt} = K m^2 \, \frac{4 k^2}{r^4} = \frac{4
\, K \, E^4}{m^2 k^2} \, . \label{power}\ee
This last equality uses the nonrelativistic expression which
relates the energy and radius of a circular orbit: $E = - 2
\,mk/(2 \, r) = -mk/r$.

The time taken for the radius to change from $r_i$ to $r_f$ is
therefore
\be \label{lifetime}
    \tau = - \int_{E_i}^{E_f} \frac{dE}{P} = \frac{m^2 k^2}{4 \, K}
    \int_{E_f}^{E_i} \frac{dE}{E^4}  = \frac{(r_i^3 - r_f^3)}{12\,mk\, K} \,
    \, .
\ee
Taking $r_f \sim \ls \ll r_i$ and comparing $\tau$ with the
initial orbital period, $T = 2 \pi \, \sqrt{r_i^3/2k}$ gives an
estimate of the number of orbits traversed before decay:
\be
    {\cal N} = \frac{\tau}{T} = \frac{r_i^{3/2}}{12\pi \sqrt{2k}\,m  K} \,
    \, .
\ee

To see how these results scale with the microscopic quantities, we
may use the D6-brane results $m = T_6 V_6 = V_6/((2 \pi)^6 \,
\ls^7 g_s)$ and $k = c_6 g_s N \ls = g_s N \ls/2$ and $K =
\kappa_{10}^2/(8 \pi \, V_6)$ with $\kappa_{10}^2 = (2 \pi)^6 \pi
g_s^2 \ls^8$. Using these results the number of orbits before
decay becomes
\be \label{Norbit}
    {\cal N} = \frac{2}{3 \pi} \, \left( \frac{r_i}{\ls} \right)^{3/2} \,
     \frac{1}{\left(g_s^3 \, N \right)^{1/2}}
    \, .
\ee

This dependence is relatively easy to understand.
\begin{itemize}
\item As already discussed, $V_6$ does not appear in
eq.~\pref{Norbit} because $P$ and $E$ are both proportional to the
brane volume.
\item The string coupling dependence arises from several sources. First
notice that the orbital energy does not depend on $g_s$ because
the contribution cancels between the interaction constant $k$ and
the tension $T_p$. For a similar reason $g_s$ drops out of the
combination $m^2 \kappa^2$ in the radiation rate. The entire
contribution of $g_s$ to the radiation rate therefore arises
through the acceleration, which vanishes if $g_s \to 0$. Taken all
together this gives a $1/g_s^2$ dependence to the lifetime, which
combines with the $g_s^{1/2}$ arising from the $k^{1/2}$ in the
orbital period, to make ${\cal N} \propto g_s^{-3/2}$.
\item The $N$ factor is the relative charge of the set of branes
to the antibrane. The bigger is $N$, the stronger is the
acceleration of the probe brane and so the shorter is the
branonium lifetime.
\end{itemize}

We see that the branonium lifetime is maximized the smaller we
take $g_s^3 \, N$ and the larger we take the orbit to be. As
discussed in previous sections, if we wish to use the full
nonlinear field equations rather than just their linearized form,
consistency requires $g_s \, N \gg 1$, and so a long-lived orbit
would require $g_s$ and $r/\ls$ to be chosen that much larger.

Of course if the brane were in the quantum limit, this orbital
decay instability could be prevented just as is done for
electromagnetic decay in atoms. Unfortunately this mechanism is
not available to us for branes in the limit of large $V_p$, since
for these the stabilization occurs at distances of order the Bohr
radius from the source branes. As we have seen earlier, in the
decompactification limit the stabilizing bound states lie beyond
the domain of validity of our approximations, since their size is
smaller than the string length. It might nonetheless be of
interest in some contexts, such as in the very early universe if
we assume that all dimensions start with a size close to the
string scale.

\subsection{Radiation into Brane Modes}
We have seen that probe brane orbits decay due to radiation into
bulk modes, and but this radiation could also be accompanied by
radiation onto the brane. We next perform an estimate of this
rate, along the lines of the one done earlier for bulk radiation,
with the conclusion that radiation into brane modes is subdominant
to bulk radiation.

We use for these purposes the coupling of the brane-bound gauge
field, given by the (String-Frame) Born-Infeld action:
\be
  S_{BI} = -T_p \int d^d \xi \; e^{-\phi} \sqrt{-
    \det(\hgamma_{MN}) \, \left[1 + \ls^4 \hgamma^{\mu\lambda}
    \hgamma^{\nu\rho} \cF_{\mu \nu} \cF_{\lambda\rho} + \cdots \right] }
    \, ,
\ee
where $\hgamma_{\mu\nu} = \hat{g}_{MN} \partial_\mu x^M
\partial_\nu x^N$ is the brane's induced metric and $\cF_{\mu\nu}$
is the gauge field strength. We see that the relevant coupling is
in this case $\kappa_B = T_p \,  \ls^4 \, e^{-\phi_b}$, where
$\phi_b = \phi(y_b)$ is the dilaton field evaluated at the
position of the brane.

Although radiation into the bulk requires non-vanishing
acceleration, this is not sufficient to produce radiation into
brane modes. This is because brane radiation arises due to the
time dependence of the fields which probe-brane-bound observers
see as a result of their motion through the fields of source
branes (much as occurs in mirage cosmology). Since this
time-dependence only arises if the interbrane distance, $r$,
changes with time, brane radiation does {\it not} occur for
circular brane orbits.

The role of $a_b^2$ in the bulk radiation rate is in this case
therefore played by $(dh/dt)^2 \sim (2k \,\dot{r}/r^2)^2 \sim (2k
\, \Omega \, e/r)^2$, where $e$ is the orbital eccentricity (which
we take to be small). Keeping in mind that translation invariance
requires the power radiated to be proportional to $V_p$ and that
the particles be produced in pairs --- with opposite momenta to
conserve momentum --- we obtain the following estimate for the
power radiated per unit spatial volume
\be
    \frac{P_p}{V_p} \sim 2\pi \, \kappa_B^2 \,
    \left(\frac{dh}{dt}\right)^2\
    \int \delta(E_1 + E_2 - \Omega) \, \delta^{p}(k_1 + k_2) \,
    \frac{d^{p}k_1 \, d^{p}k_2}{(2 \pi)^{2p} \, E^{2p-7}}
    \sim \frac{\kappa_B^2 \, \dot{h}^2 \, \Omega^{6-p}}{(2\pi)^{{\cal
    P}}} \, ,
\ee
where the power of energy is included on dimensional grounds and
${\cal P} = 2p -1 - \left[{p}/{2} \right]$.

Choosing 3-branes, and using $\kappa_B = T_4 \, \ls^4 \,
e^{-\phi_b} = 1/g_b$, where $g_b = g_s e^{\phi_b}$ is the string
coupling at the position of the brane, gives $[p/2] = 1$ and so
\be \label{BraneRad}
    \frac{P_3}{V_3} \sim \frac{\Omega^3 \, \dot{h}^2 }{g_b^2
    (2\pi)^{4}} \sim \frac{e^2 \, \Omega^5 }{g_b^2 (2 \pi)^4}
    \left( \frac{2k}{r} \right)^2
    \sim \frac{e^2}{(2\pi)^4 \, g_s^2 \, r^5} \, \left(
    \frac{2k}{r} \right)^{9/2} \sim \frac{e^2}{(2\pi)^4 \, g_s^2 \, r^5} \, \left(
    \frac{g_s N \, \ls}{r} \right)^{9/2} \, .
\ee
Here we use the large-$r$ results $\Omega = \sqrt{2k/r^3}$, $g_b =
g_s [1 - (2k \kappa/r) + \cdots]$, as well as the relation $2 \, k
\sim g_s N \, \ls$. The dependence of this estimate on $\Omega$,
$e$, and $k/r$ is borne out by a toy calculation of the rate of
change of the vacuum energy density on the brane which is given in
the appendix.

Several conclusions may be drawn from this estimate:
\begin{itemize}
\item Comparing the power radiated into the bulk \pref{BulkRad}
and \pref{BraneRad} gives the ratio of power radiated into the
bulk and to the brane. For D3-branes moving in 3 transverse
dimensions this ratio becomes:
\be
    \frac{P_{\rm brane}}{P_{\rm bulk}} \sim \frac{ e^2 \,
    \Omega^3}{g_s^2 (2\pi)^{3} \kappa_7^2 \, T_4^2\,
    a_b^2}\left(\frac{dh}{dt} \right)^2
    \sim \frac{e^2 \, \ls^3}{g_s^2 (2 \pi)^3 \, r^3} \left( \frac{2k}{r}
    \right)^{5/2}
    \sim \frac{e^2 (g_s N)^{5/2}}{(2\pi)^3 g_s^2} \, \left(
    \frac{\ls}{r} \right)^{11/2}
    \, .
\ee
This shows that the greater part of the emitted radiation goes
into the bulk.
\item Although most of the radiation goes into the bulk, particles
{\it are} produced on the branes, and this has possible
implications for cosmology. Provided the number of source branes
differs from the number of probe branes, some branes survive even
after the orbit eventually decays. If, for instance, we imagine
ourselves being trapped on one of these remaining branes, these
particles can equilibrate with observable particles at later
times. Depending on the details, this could provide a source of
reheating and/or baryogenesis. For a branonium-inflation scenario
these produced particles are likely to be inflated away as they
are produced, and so are unlikely to obviate the necessity for
other sources of reheating, such as due to the brane-antibrane
annihilation itself.
\end{itemize}

\section{Conclusions}
We have seen that the bound states of brane systems have very
interesting properties. The present work should be considered only
as the beginning of their exploration. Within a brane world
context we find it appealing to imagine our entire universe as
being, albeit briefly, a member of a `solar system' of universes.

Our analysis has included only branes and antibranes which share
the same dimensionality and are parallel, which makes a closer
analogy with positronium than with a stable atom. It is
straightforward to generalise our discussion to a system of branes
having different dimensionality and/or branes at angles, for which
the effective interactions can still be attractive. We would
expect the behaviour and instability to annihilation of these
systems to be very similar, although a detailed analysis is left
for the future. Similarly one might extend this idea to possible
bound states between branes and orientifold planes. Other systems,
including lifts to 11 dimensions and/or $T$- and $S$-\,dual
versions might also be considered.

Many open questions remain. One is a more fundamental
understanding of why D-brane/antibrane orbits do not precess,
having the same shape for fully relativistic systems as they have
in the non-relativistic limit. There may exist a mapping between
the two Lagrangians in a similar way as recently found in
ref.~\cite{bellucci} for the case of a massive particle near the
horizon of an extreme Reisner-Nordstr\"om black hole. Notice that
Bertrand's theorem \cite{g}\ states that the only systems giving
rise to closed orbits in the central force problem are the $1/r$
potential and the harmonic oscillator. The D-brane systems we
consider here are not counterexamples to this theorem because of
the canonical kinetic terms which the proof of Bertrand's theorem
assumes. It would be interesting to see how other classical
results, such as the virial theorem, the three- (or more) body
problem or chaotic orbits, generalize in the present instance.

Our quantum treatment of the branonium system was very basic. A
better understanding of how to properly describe the system at the
quantum level and the physical interpretation of the corresponding
wave function would be desirable, perhaps with implications to
quantum cosmology. It would also be interesting to know in more
detail the back reaction of the probe brane on the bulk metric,
both to better understand the $N=1$ brane/antibrane problem and to
have an explicit microscopic derivation of the long-distance
fields which are of interest for cosmological applications.
Because the branes behave as point masses in the transverse
dimensions it is clear that for 3 or more transverse dimensions
the bulk gravitational (and other) fields in the extra dimensions
can be made negligible simply by keeping them well separated.
Experience with co-dimension 1 branes can be misleading in this
case, because the bulk fields which these give rise to can instead
be homogeneous spacetimes like de Sitter or anti-de Sitter space.

Finally, the applications to cosmology are potentially many.
Inflation can be analyzed in a way similar to ref.~\cite{OurBI}.
In this case the system has several novel features. For instance
it provides an unusual slow roll because it is the angular
variable, $\theta$, which rolls even though the potential is
exactly flat in this direction. Nonetheless, the constant
inflationary energy density can be expected to evolve as the
orbital radius changes due to energy loss into bulk radiation.
This changes the analysis in an interesting way, since $\dot r$ is
not related to $dV/dr$ in the usual expression. Furthermore, the
fact that the orbit is unstable and decays is likely to make the
brane motion appear as a very unusual `fluid' from the
cosmological point of view. As for other forms of brane/antibrane
inflation, the end of inflation would be occur once the antibrane
falls into the source branes and annihilates, providing the usual
extremely rich realization of hybrid inflation.

The orbital motion of the probe brane might also provide other
novel cosmologies, such as a particularly interesting example of a
mirage cosmology. For elliptical orbits the periodic increase and
decrease of the interbrane separation gives rise to a cyclic
cosmology for the trapped brane observers. Because the small size
of the extra dimensions implies the period of these cycles is
microscopically short for standard compactifications, this is
likely to be relevant only for very early cosmology. The particle
production which is produced by the time dependence of the induced
fields on the branes in these cosmologies may have implications,
perhaps providing a geometric realization of the Affleck-Dine
mechanism for baryogenesis. Because these speculations take us
considerably beyond the scope of the present paper, we devote a
companion article \cite{branonium2} to a more detailed discussion
of these cosmological issues.

\section{Appendix: Quantum Energy on the Brane}
We here briefly record a toy calculation of the time dependence of
the quantum stress energy density on the brane which arises due to
the time dependence of the induced brane metric.

Our toy model consists of a conformally coupled scalar, in 4
spacetime dimensions, defined by the action
\be
    \frac{{\cal L}_{\rm toy}}{\sqrt{-g}} = - \, \sfrac12 \,
    (\partial \varphi)^2 - \sfrac16 \, R \varphi^2 \, .
\ee
We evaluate the quantum stress energy tensor for this scalar using
a time-dependent but conformally-flat metric
\be \label{confmet}
    ds^2 = h^{-1/2} (- dt^2 + d\xi^2) \, ,
\ee
where $h = 1 + k/r$ and $r = r_0 \, \Bigl[1 + e \, \cos(\Omega t)
\Bigr]$. Here $e$ and $\Omega$ represent the eccentricity and the
angular frequency of the radial oscillations for a slightly
eccentric elliptical orbit.

Now, for any theory which is conformally trivial, the expectation
value of the stress energy tensor, $T_{\mu\nu}$, is determined by
integrating the conformal anomaly, leading to the expression
\cite{BD}
\begin{eqnarray}\label{eom1}
    \langle \, T_{\mu\nu}\rangle  &=& \frac{1}{1440\pi^{2}}
    \left[ aR_{;\mu\nu} - a g_{\mu\nu} { \Box} R + \left( a - \frac{b}{3}
    \right) R R_{\mu\nu} \right. \nonumber \\
    &&\qquad + \left. \left( - \frac{a}{4} + \frac{b}{8} \right)
    g_{\mu\nu} R^{2} + \frac{b}{2}
    R_{\mu\alpha} R_{\nu}^{\alpha} - \frac{b}{4} g_{\mu\nu}
    R_{\alpha\beta}R^{\alpha\beta} \right] \, .
\end{eqnarray}
Here the constants $a$ and $b$ depend on the spin of the field in
question. For a conformal scalar fields we have $a=- \sfrac16$ and
$b = 1$, while for a massless vector field we have, $a = -3$ and
$b=62$.

Inserting the metric, eq.~\pref{confmet}, into eq.~\pref{eom1},
making use of the relations ${\omega}= \sqrt{2k/r^{3}}$, and
$m=T_{p}V_{p}$ we find (to leading order in ${1}/{r_0}$ and $e$):
\be
    \rho = \langle \, T_{00} \rangle = \frac{{\Omega}^{4}
    k^{2} \, e^{2}}{2560\pi^{2}r_0^{2}} \Bigl[ 2a+(b-3a)
    \cos^2({\Omega}t) \Bigr] \, .
\ee
An estimate for the energy production rate on the brane is then
\be
    \frac{P}{V_3} = \frac{d\rho}{dt} = \frac{ {\Omega}^{5}
    k^{2}\, e^{2}}{1280\pi^{2}r_0^{2}} (-b+3a)
    \cos({\Omega}t) \sin({\Omega}t) \, .
\ee
Notice that this scales like $e^2 \, \Omega^5 (2k/r)^2$, just as
does our estimate in the main text for the energy deposition rate
for the brane.

\section*{Acknowledgments}
We would like to acknowledge helpful conversations with Robert
Brandenberger, Jim Cline, Atish Dabholkar, Roberto Empar\'an,
Jaume Garriga, Joaquim Gomis, Nicol\'as Grandi, David Grellscheid,
David Mateos, John March-Russell, Anupam Mazumdar, Rob Myers,
Rub\'en Portugues, Harvey Reall, Paul Townsend, Ivonne Zavala.
C.B.'s research is supported in part by grants from N.S.E.R.C.
(Canada), F.C.A.R. (Qu\'ebec) and McGill University. F.Q. is
supported by PPARC. F.Q. thanks the CERN Theory Divison for
hospitality and  R.R. thanks the Physics Department of McGill
University.

\end{document}